\documentclass[preprint,review,12pt]{elsarticle}

\usepackage{caption}
\usepackage{subcaption}

\usepackage{amssymb}
\usepackage{isotope}
\usepackage{siunitx}
\usepackage{chemformula}
\usepackage[section]{placeins}

\usepackage[T1]{fontenc}
\usepackage{listings}
\usepackage[framed,numbered]{matlab-prettifier}
\journal{Journal of Magnetic Resonance Open}

\begin{document}

\begin{frontmatter}

\title{Phase distortion-free paramagnetic NMR spectra}

\author[inst1,inst2,inst3,inst4]{Enrico Ravera\corref{cor1}}
\cortext[cor1]{Corresponding Author: ravera@cerm.unifi.it}

\affiliation[inst1]{organization={Department of Chemistry ``Ugo Schiff", University of Florence},
            addressline={Via della Lastruccia 3}, 
            city={Sesto Fiorentino},
            postcode={50019}, 
            country={Italy}}

\affiliation[inst2]{organization={Magnetic Resonance Center, University of Florence},
            addressline={Via Luigi Sacconi 6}, 
            city={Sesto Fiorentino},
            postcode={50019}, 
            country={Italy}}

\affiliation[inst3]{organization={Consorzio Interuniversitario Risonanze Magnetiche di Metalloproteine},
            addressline={Via Luigi Sacconi 6}, 
            city={Sesto Fiorentino},
            postcode={50019}, 
            country={Italy}}

\affiliation[inst4]{organization={Florence Data Science, University of Florence},
            country={Italy}}

\begin{abstract}

The NMR spectra of paramagnetic substances can feature shifts over thousands of ppm. In high magnetic field instruments, this corresponds to extreme offsets, which make it challenging or impossible to achieve uniform excitation without phase distortion. Furthermore, because of the intrinsic presence of a dead time during which spins freely evolve, a further phase distortion occurs. A decade ago, a processing approach based on the analysis of the statistics of the phase of the spectrum was proposed to denoise NMR spectra. In this manuscript it is demonstrated that this approach is applicable to obtain paramagnetic NMR spectra that are free of phase distortion, even though the quantitative information of peaks intensities is lost. This is demonstrated on the high field spectra of a prototypical nickel(II)-complex and through the analysis of simulated data.
\end{abstract}

\begin{graphicalabstract}
\hspace{0cm}\includegraphics[width=5cm]{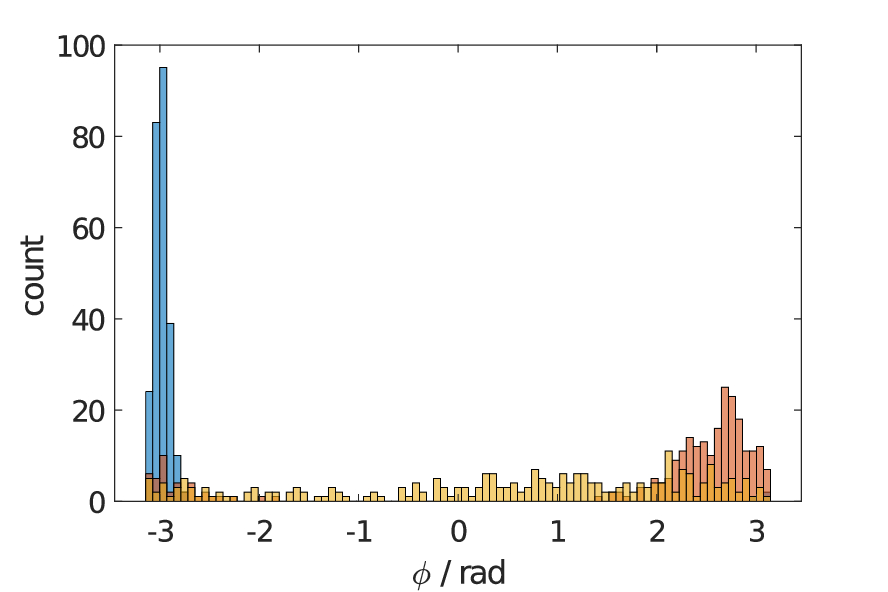}\hspace*{0cm}\includegraphics[width=5cm]{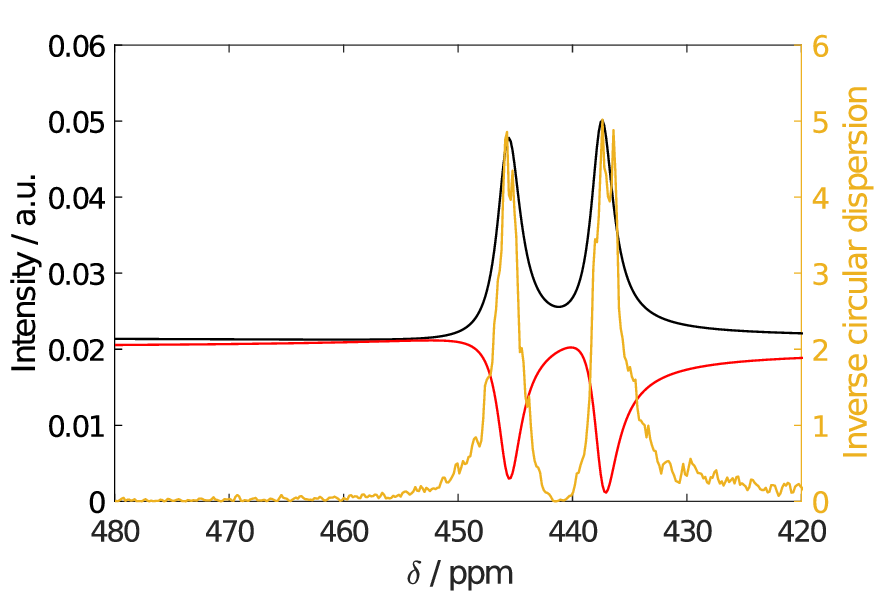}\\
Directional statistics of phases $\longrightarrow$ Phased spectrum
\end{graphicalabstract}

\begin{highlights}
\item Spectra of paramagnetic compounds have signals over a broad range of frequencies.
\item Finite pulse lengths and the dead time before acquisition start result in severe phase distortion.
\item First-order phasing is usually applied, yielding severe baseline distortion.
\item Statistical analysis of the phase across several repetition of the same experiment can yield distortionless spectra.
\end{highlights}


\end{frontmatter}


\section{Introduction}
\label{sec:introduction}
NMR of paramagnetic molecules (pNMR) has been pioneered in the sixties and early seventies by a relatively small but very active group of leading chemists \cite{la1964isotropic,bertini1969detection,holm1969applications,morishima1970nuclear,bertini1972proton,bleaney1972origin}, who saw the possibilities offered by these technique in understanding structure, dynamics and electronic properties of paramagnetic coordination compounds. In the eighties, the bioinorganic applications bloomed\cite{bertini_nmr_1986}. As the interest in the elucidation of the properties of paramagnetic compounds steadily increases, because of the applications in healthcare (MRI contrast agents \cite{aime2018relaxometry,peters2020chemical}), quantum information processing (single ion magnets, qubits \cite{ishikawa2003determination,damjanovic2013combined,damjanovic2015ligand,hiller2017ligand,parker2020ligand,gigli2021nmr}) and biomedicine (metalloproteins, \cite{knight2013magic,ravera2018nmr,bertarello2018solid,ciambellotti2019structural,silva2019metal,piccioli2020paramagnetic}), the applications of pNMR are increasing as well, because of the unique ability of NMR to detect structural and dynamical features at the atomic level \cite{ravera2019methodological}.\\  
The possibilities offered by NMR have been dramatically boosted by the advent of modern instrumentation and by the accessibility and quality of Quantum Chemical methods for the calculation of pNMR observables \cite{bertarello2020picometer,ravera2021quantum,anie.202101149,gade2021observability}. 
\\
However, the experiments remain challenging, expecially those that are aimed at detecting very far shifted resonances \cite{ott2018taking,gade2021observability} and in very high magnetic fields \cite{ravera2021quantum}. One aspect in particular can be rather tedious to resolve: the phase distortion, which will be described in the next section (\ref{subsec:phasedist}). In brief, the signals have larger 1st order phase distorsion the larger is their offset from the carrier frequency. To phase the spectra corresponds to introducing a rolling in the baseline, that must be dealt with \textit{a posteriori}. In this manuscript we explore the application of a very ingenious processing method proposed by Takegoshi and co-workers \cite{fukazawa2010phase,fukazawa2011post} to resolve the issue of phase distorsion across a pNMR spectrum.
\subsection{Phase distorsion in pNMR}
\label{subsec:phasedist}
\begin{figure}
    \centering
    \begin{subfigure}[t]{0.49\textwidth}
        \centering
        \includegraphics[width=0.95\linewidth]{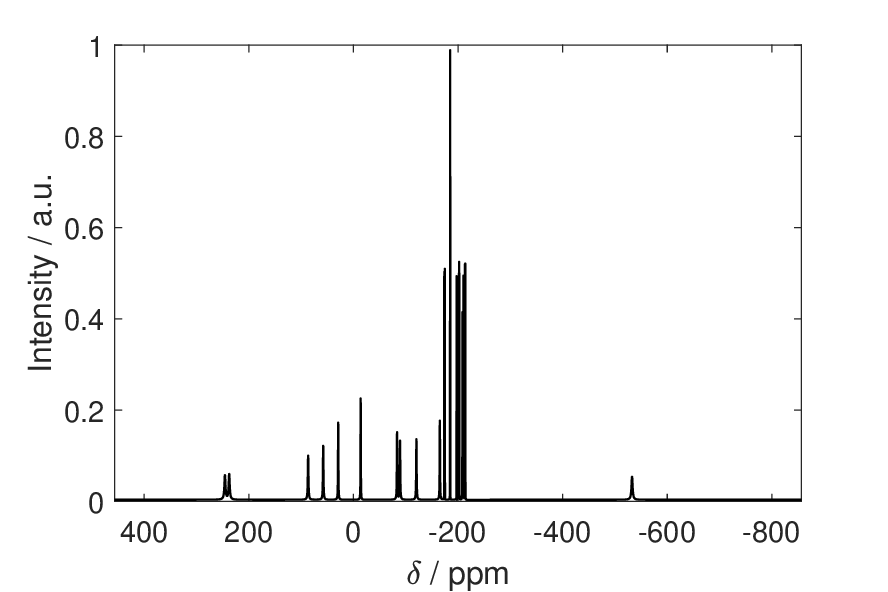} 
        \caption{Ideal spectrum} \label{fig:fig1a}
    \end{subfigure}
    \hfill
    \begin{subfigure}[t]{0.49\textwidth}
        \centering
        \includegraphics[width=0.95\linewidth]{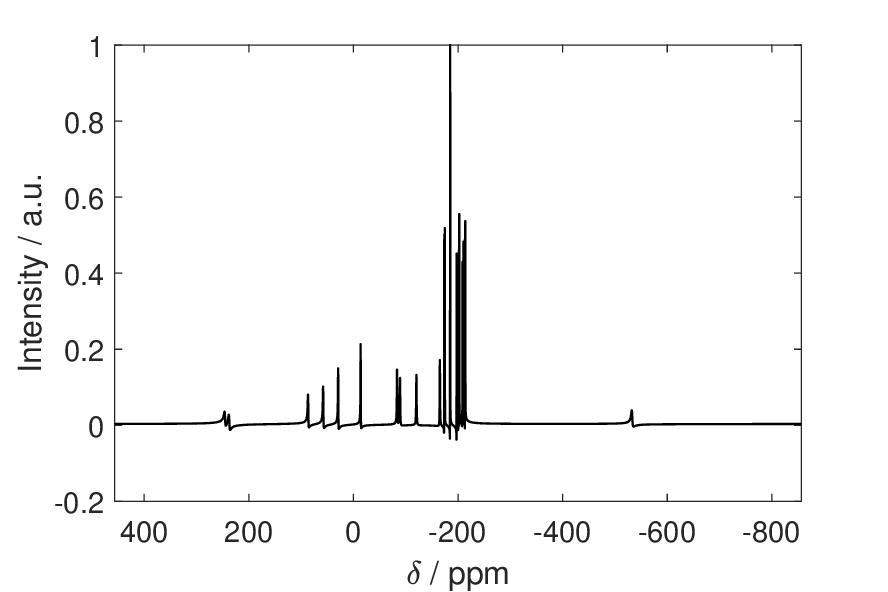} 
        \caption{Effect of finite length of the pulse} \label{fig:fig1b}
    \end{subfigure}

    \begin{subfigure}[t]{0.49\textwidth}
        \centering
        \includegraphics[width=0.95\linewidth]{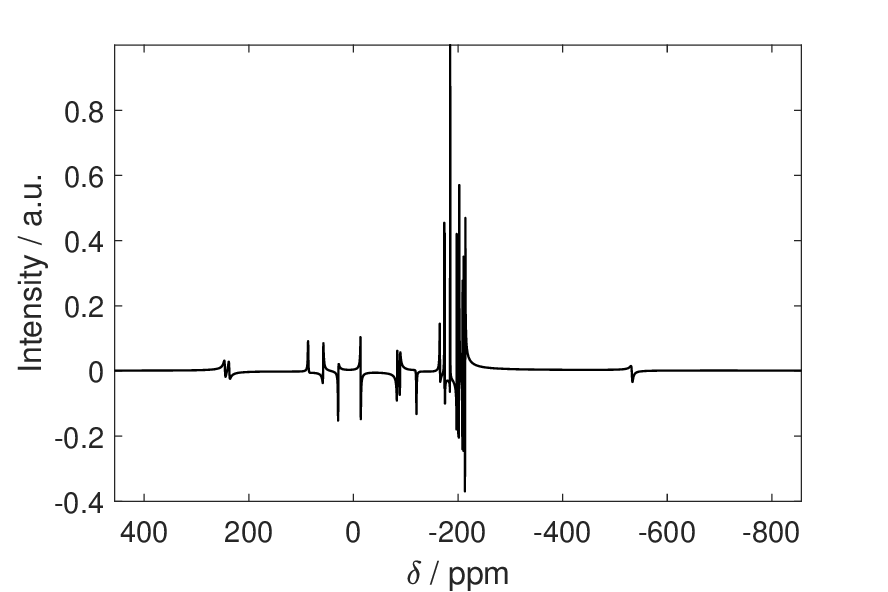} 
        \caption{Effect of dead time} \label{fig:fig1c}
    \end{subfigure}
    \hfill
    \begin{subfigure}[t]{0.49\textwidth}
        \centering
        \includegraphics[width=0.95\linewidth]{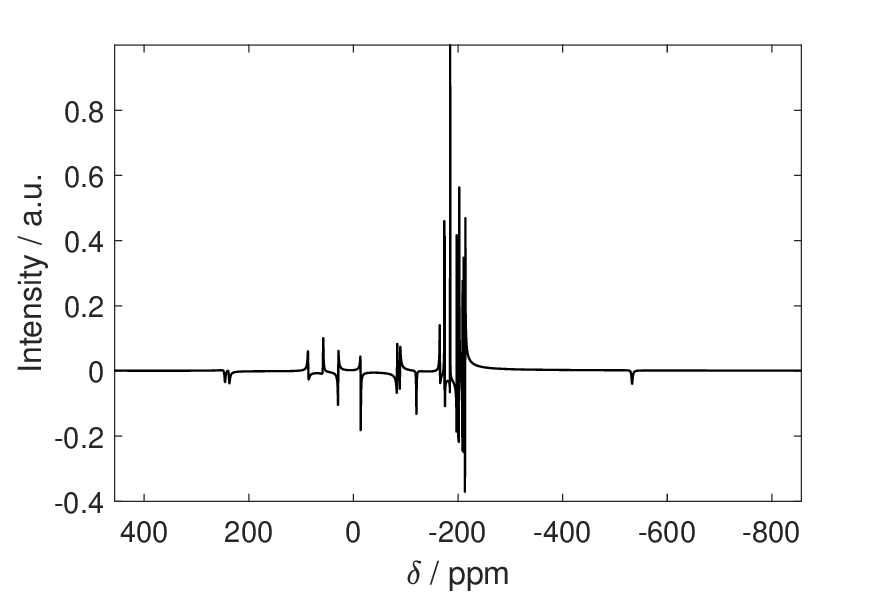} 
        \caption{Distorted spectrum} \label{fig:fig1d}
    \end{subfigure}
    
    \begin{subfigure}[t]{0.49\textwidth}
        \centering
        \includegraphics[width=0.95\linewidth]{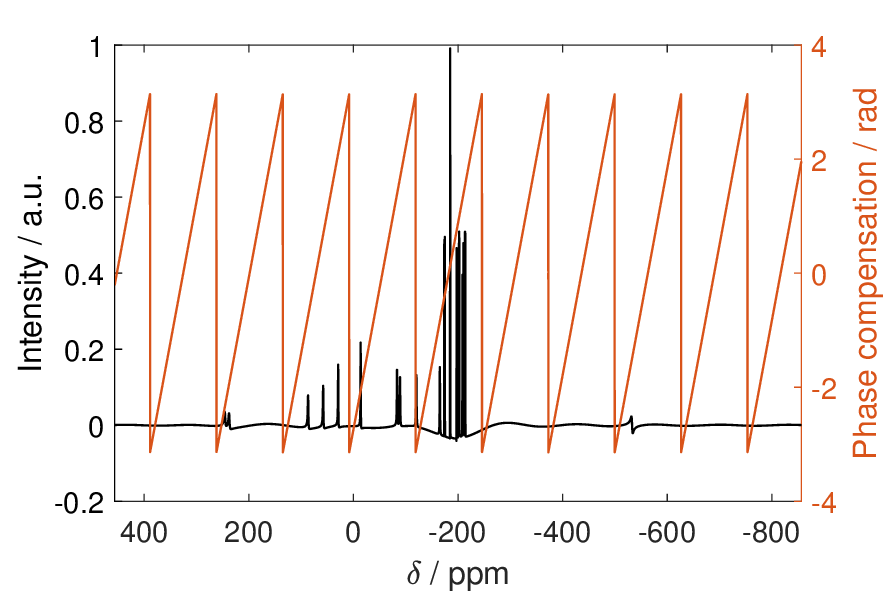} 
        \caption{Approximately phased spectrum with baseline distortion} \label{fig:fig1e}
    \end{subfigure}
    \caption{Simulated spectra of Ni-SAL-HDPT\cite{sacconi1966high} at \SI{22.4}{T} (\SI{950}{MHz} \isotope[1]{H} Larmor frequency) assuming homogeneous excitation and no dead time \subref{fig:fig1a}, including only the effect of a finite pulse (excitation pulse of \SI{1}{\us} with a nutation frequency of \SI{35.7}{kHz}, corresponding to a \ang{90} pulse of \SI{7}{\us}) \subref{fig:fig1b}, assuming a dead time of \SI{8}{\us} \subref{fig:fig1c} and combining dead time and pulse imperfection \subref{fig:fig1d}. Panel \subref{fig:fig1e} shows the baseline distortion effect of phasing the spectrum (in black) and the required profile of phase compensation (in red). The spectrum is assumed to be free of probehead background.}     \label{fig:fig1}
\end{figure}

In a pNMR spectrum, a 1st order phase distorsion is usually present because of two factors, that are intrinsic of the experiment. They will both be exemplified in a simulated spectrum of a model paramagnetic complex (Ni-SAL-HDPT) \cite{sacconi1966high,ravera2021quantum}, which features shifts over about 1000 ppm range. The idealized spectrum is depicted in Figure \ref{fig:fig1}, panel \subref{fig:fig1a}. One contribution to the phase distortion comes from the fact that the effect of a pulse of finite length and finite nutation frequency, even if perfectly rectangular, yields an excitation profile in intensity and phase that changes with the offset of the signal from the carrier frequency (Figure \ref{fig:fig1}, panel \subref{fig:fig1b}). The other contribution to the phase distortion is the presence of a dead time between the end of the pulse and the opening of the receiver, during which the signals evolve, and this translates into an additional phase contribution (Figure \ref{fig:fig1}, panel \subref{fig:fig1c}). The evolution of the signals during the dead time can be compensated through echo detection for relatively small spectral windows \cite{bertini2016nmr}, but it becomes unpractical for the larger ones because of the need of ultrashort broadband refocusing pulses \cite{asami2018ultrashort} for obtaining a detectable echo. The overall spectrum results extensively distorted (Figure \ref{fig:fig1}, panel \subref{fig:fig1d}). A phase correction - if at all possible - results in a severe baseline distortion (Figure \ref{fig:fig1}, panel \subref{fig:fig1e}). The latter can only in part mitigated through backwards linear prediction \cite{bertini2016nmr}. Magnitude processing significantly alters the lineshape of the peaks (Figure \ref{fig:figSmagn}): this is related to the broad "wings" of the Kramers-Kroning related function in the imaginary channel (dispersion) of a phased peak in the real channel (absorption).

\subsection{The phase covariance method}
\label{subsec:phasecov}
\begin{figure}[htbp]
  \centering
  \begin{subfigure}{0.56\textwidth}
        \includegraphics[width=0.95\linewidth]{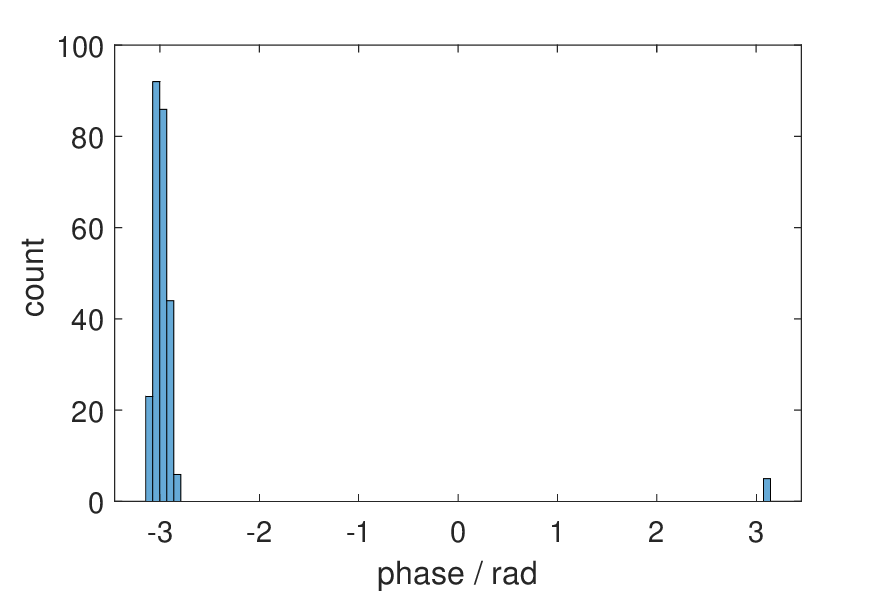} 
    \caption{Signal at 186.2 ppm}
    \label{fig:fig2a}
  \end{subfigure}
   \begin{subfigure}{0.56\textwidth}
        \includegraphics[width=0.95\linewidth]{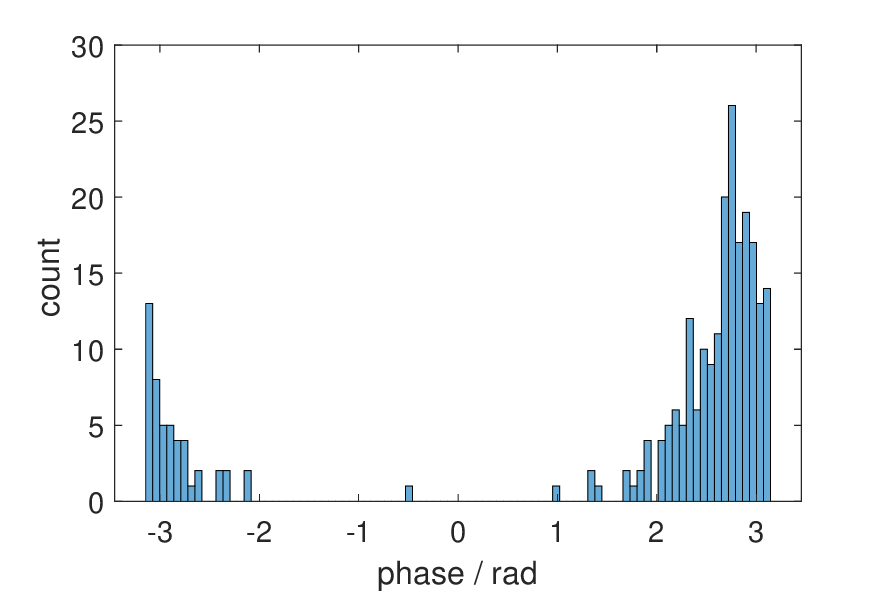} 
    \caption{Signal at 437.5 ppm}
    \label{fig:fig2b}
  \end{subfigure}
  \begin{subfigure}{0.56\textwidth}
        \includegraphics[width=0.95\linewidth]{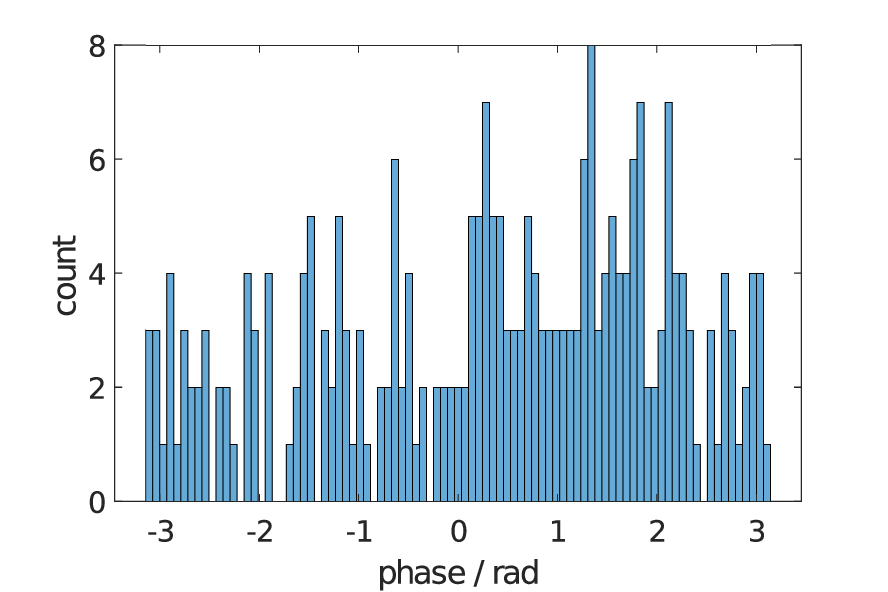} 
    \caption{Noise at 153 ppm}
    \label{fig:fig2c}
  \end{subfigure}
  \caption{Histogram of phase angles in points of the spectrum corresponding to two signals of different intensities (\subref{fig:fig2a}, higher intensity, \subref{fig:fig2b}, lower intensity) and to noise \subref{fig:fig2c}}\label{fig:fig2}
\end{figure}

In a brilliant paper that appeared in 2010 \cite{fukazawa2010phase}, Fukazawa and Takegoshi ingeniously noted that the phase of the signal must be related to the phase of the last pulse in the pulse sequence, and that this information could be used to discriminate between the signal and the noise. The method was applied to denoise the spectra of a low-concentration mixture of l-alanine (3 wt\%) and glycine (1 wt\%) in KBr powder, as well as to discriminate between proper peaks and artifacts. One year later \cite{fukazawa2011post}, the same authors with Takeda also discussed an approach (from here on referred to as FTT) based on the acquisition of multiple spectra (with a complete phase cycle) and the application of descriptive statistics on each point of the spectrum to discriminate the signal and the noise. The method was applied on the same system. In a nutshell, the method works as follows: the complex spectrum is separated in magnitude and phase, and the standard deviation of the phase is evaluated. 
If a point in the spectrum contains signal, the values of phase will be distributed around a given value, and the spread of the distribution will be lower the higher the intensity of the signal. On the contrary, if a point in the spectrum contains noise, the values of the phase will be distributed homogeneously over the circle. This is exemplified in figure \ref{fig:fig2}.
The application of the FTT method yields a ``denoised spectrum'' that reflects the degree of certainty (or uncertainty) of the phase in each particular point of the spectrum. This spectrum faithfully reflects the frequencies of the signals in the original spectra but yields slightly perturbed intensities. The original method was implemented for a particular application in solid-state NMR, but it can be expected that it could be applicable in any broadband or wideline NMR application.\\
It is here argued that the idea of evaluating the statistics of the phase of the signals in the spectrum can be used to obtain a in-phase spectrum regardless of the offset of the signals from the carrier, as long as they can be excited to detection.
\section{Materials and methods}
\subsection{The sample}
The Ni-SAL-HDPT sample has been prepared and purified as described elsewhere \cite{sacconi1966high,ravera2021quantum}, dissolved in \ch{CDCl_3} and transferred to a \SI{3}{mm} tube.\\
\subsection{Experimental details}
\label{expdetails}
Spectra were acquired on a Bruker Avance III spectrometer operating at \SI{400}{MHz} \isotope[1]{H} Larmor frequency (\SI{9.4}{T}) using a \SI{5}{mm}, \isotope[1]{H}-selective probe dedicated to paramagnetic systems (the nutation frequency of the hard pulse is ca. \SI{90}{kHz}), and on a Bruker Avance III spectrometer operating at \SI{950}{MHz} \isotope[1]{H} Larmor frequency (\SI{9.4}{T}) using a triple resonance TCI cryo-probehead (the nutation frequency of the hard pulse is ca. 37 kHz. All spectra were acquired with the standard pulse-acquire sequence from Bruker library (zg) accumulating the 8 scans required for the phase cycling and repeating the experiment 256 times.\\
\subsection{Computational details}
The spectra were simulated using a purpose-written MATLAB script. The shifts were calculated as described in \cite{ravera2021quantum}, and relaxation parameters were estimated from the electronic properties of nickel(II) complexes as described in chapter 4 of \cite{bertini2016nmr}. Phase distortion due to finite pulse length is calculated as described in \cite{gregory2009effects}.\\
Experimental spectra were imported in MATLAB using the GNAT tool \cite{castanar2018gnat}.\\ 
The statistical analysis of the phase of the spectra was also performed using a purpose-written MATLAB script, based on the functions of the CircStat toolbox \cite{berens2009circstat}. There are two differences with respect to the original FTT approach: \begin{enumerate}
    \item the information arising from the original (averaged) spectrum is discarded;
    \item the circular dispersion (see definition below) instead of the standard deviation is calculated.
\end{enumerate}  
This choice is motivated by the observation that very broad lines are slightly distorted using the standard deviation (this behavior is currently being investigated, see an example of the effect on the spectrum in figure \ref{fig:figS1}). This method will be referred to as ``modified FTT'' or mFTT.
Circular dispersion is calculated as:
\begin{equation}
\label{eq:circulardisp}
    \overline{\delta} = \dfrac{1-\overline{R_2}}{2R^2}
\end{equation}
where $\overline{R_2} = \left|\dfrac{1}{N} \sum_i^N e^{i2\theta}\right|$ is the length of the second moments and $R = \left|\dfrac{1}{N} \sum_i^N e^{i\theta}\right|$ is the population length.
\\The instructions and the code for running the mFTT analysis are given in the supplementary material.

\section{Results and Discussion}
\subsection{Simulated Tests}
\label{subsec:synthtest}
\begin{figure}
    \centering
    \begin{subfigure}[t]{0.70\textwidth}
        \centering
        \includegraphics[width=0.95\linewidth]{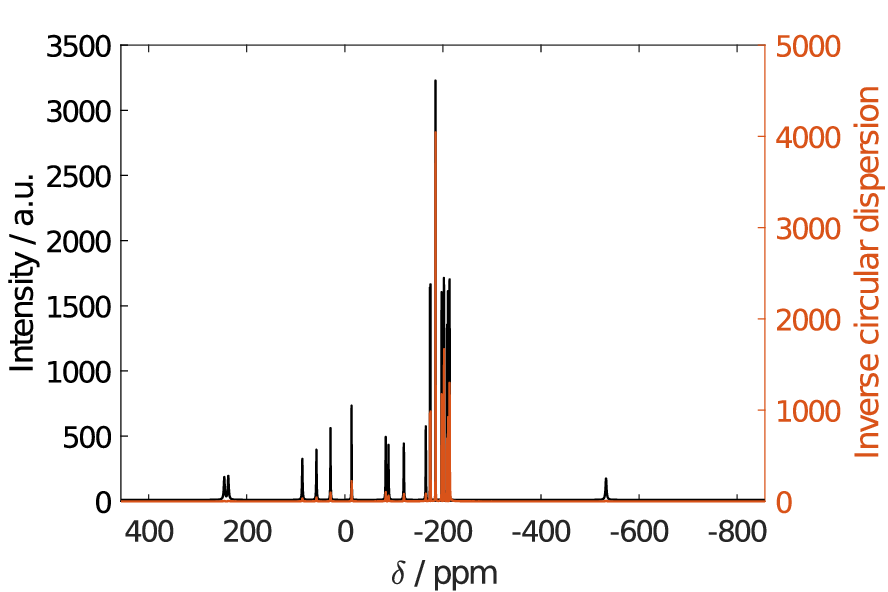} 
        \caption{Full spectrum} \label{fig:fig3a}
    \end{subfigure}

    \begin{subfigure}[t]{0.49\textwidth}
        \centering
        \includegraphics[width=0.95\linewidth]{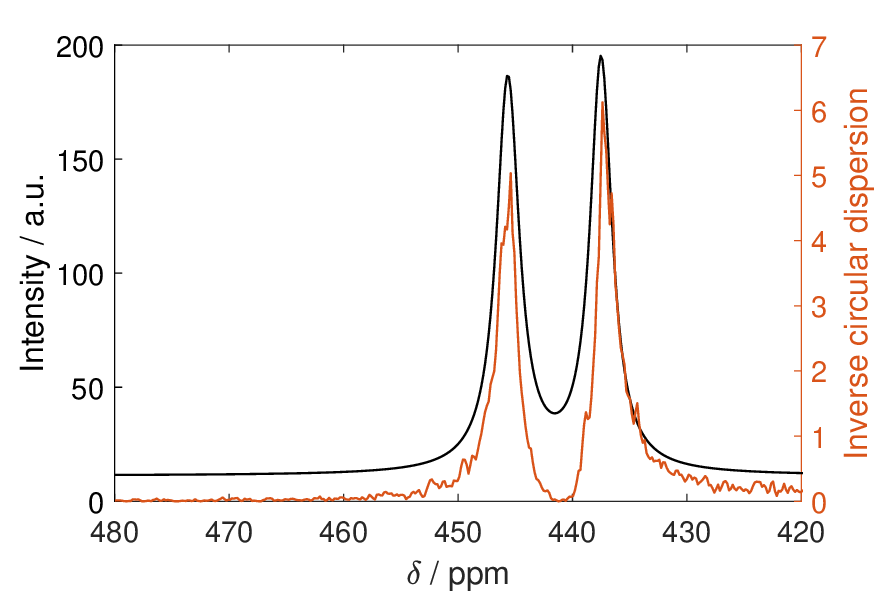} 
        \caption{Detail: 480 to 420 ppm} \label{fig:fig3b}
    \end{subfigure}
    \hfill
    \begin{subfigure}[t]{0.49\textwidth}
        \centering
        \includegraphics[width=0.95\linewidth]{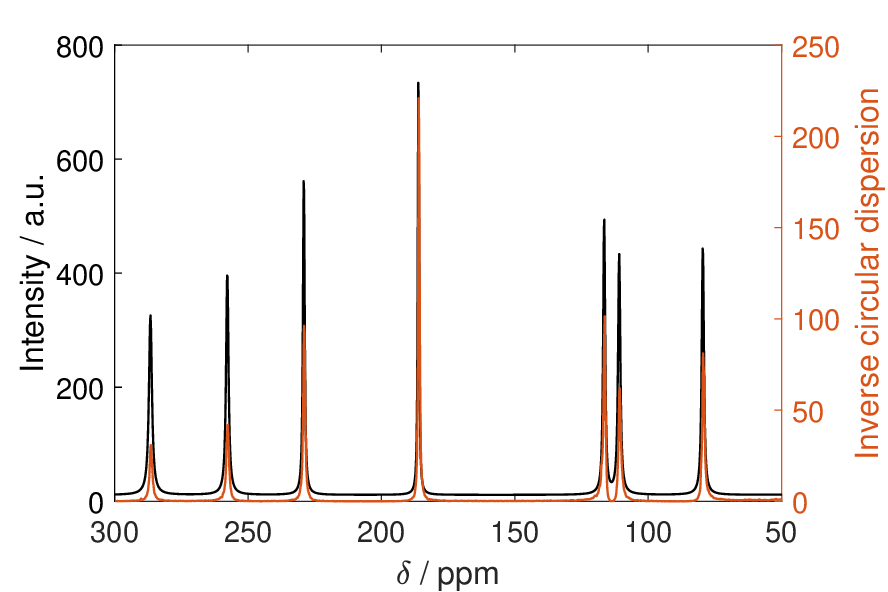} 
        \caption{Detail: 300 to 50 ppm} \label{fig:fig3c}
    \end{subfigure}
    \begin{subfigure}[t]{0.49\textwidth}
        \centering
        \includegraphics[width=0.95\linewidth]{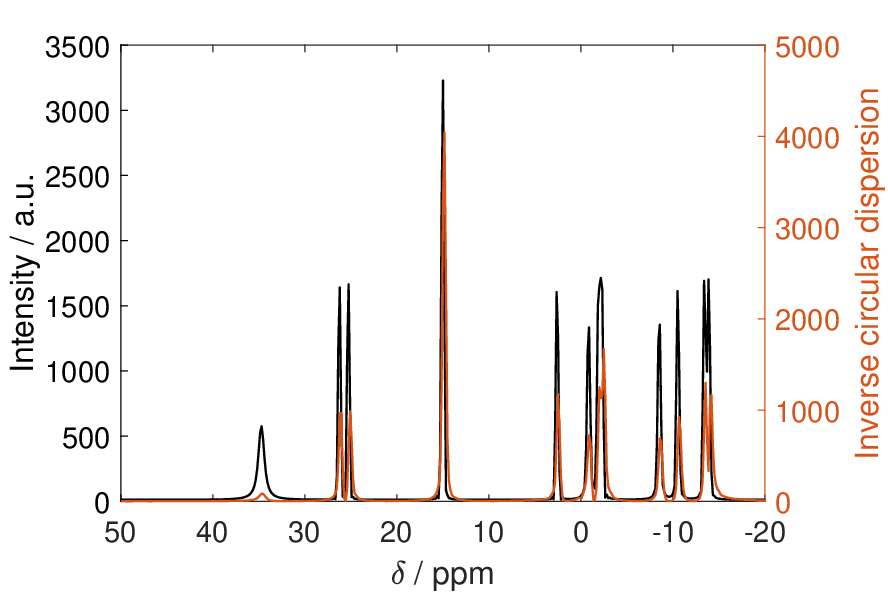} 
        \caption{Detail: 50 to -20 ppm} \label{fig:fig3d}
    \end{subfigure}
    \hfill
    \begin{subfigure}[t]{0.49\textwidth}
        \centering
        \includegraphics[width=0.95\linewidth]{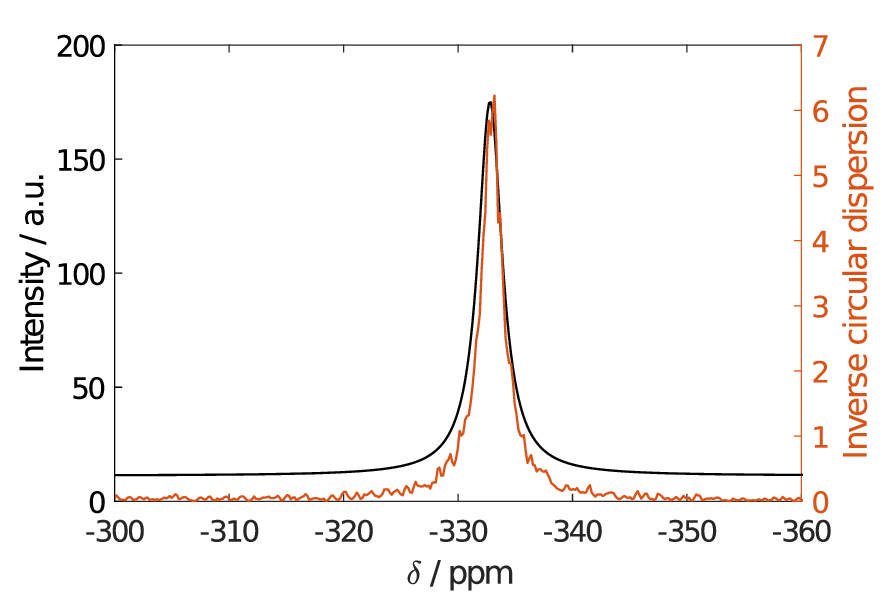} 
        \caption{Detail: -300 to -360 ppm} \label{fig:fig3e}
    \end{subfigure}

    \caption{Simulated spectra of Ni-SAL-HDPT\cite{sacconi1966high} at \SI{22.4}{T} (\SI{950}{MHz} \isotope[1]{H} Larmor frequency) assuming homogeneous excitation and no dead time \subref{fig:fig3a} in black, and the reconstructed spectrum devoid of phase distortion in red. Panels \subref{fig:fig3b}-\subref{fig:fig3e} show different details of the full spectrum.}     \label{fig:fig3}
\end{figure}

Initially, the mFTT method has been tested on simulated data representing the spectrum of Ni-SAL-HDPT acquired at \SI{22.4}{T} (\SI{950}{MHz} \isotope[1]{H} Larmor frequency), with an excitation pulse of \SI{1}{\us} at a nutation frequency of \SI{35.7}{kHz}, corresponding to a flip angle of \ang{12.9}. The spectra are shown in figure \ref{fig:fig1}. It is assumed that the probehead (and the electronics in general) behave ideally in terms of response. Normally-distributed random noise with a standard deviation of 20\% of the signal is then added to the simulated FID over 256 repetitions.\\
The circular dispersion is evaluated over the 256 transients using equation \eqref{eq:circulardisp}, and the results are plotted in figure \ref{fig:fig3} superimposed to the distortionless spectrum of figure \ref{fig:fig1}, panel \subref{fig:fig1a}.

\begin{figure}
    \centering
    \begin{subfigure}[b]{0.49\linewidth}        
        \centering
        \includegraphics[width=\linewidth]{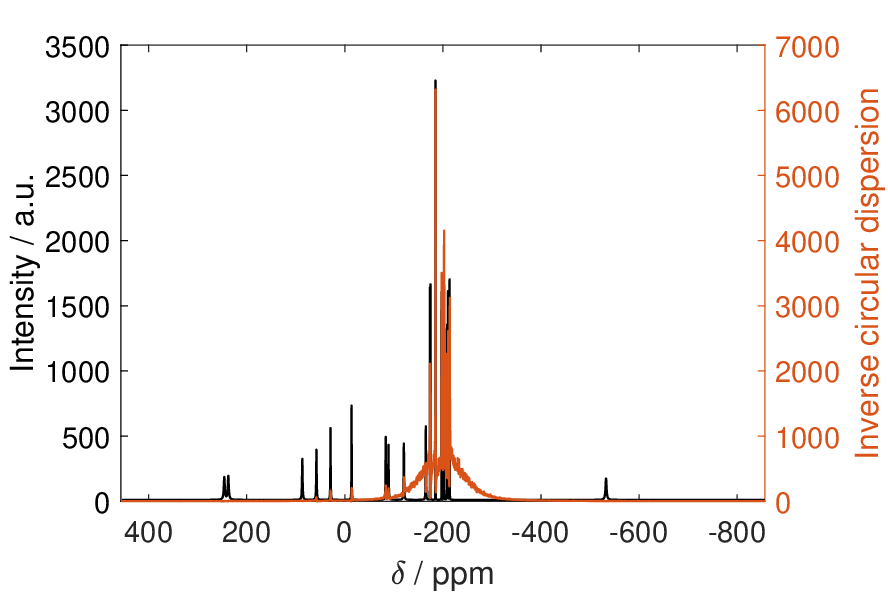}
        \caption{Full spectrum}
        \label{fig:fig4a}
    \end{subfigure}
    \begin{subfigure}[b]{0.49\linewidth}        
        \centering
        \includegraphics[width=\linewidth]{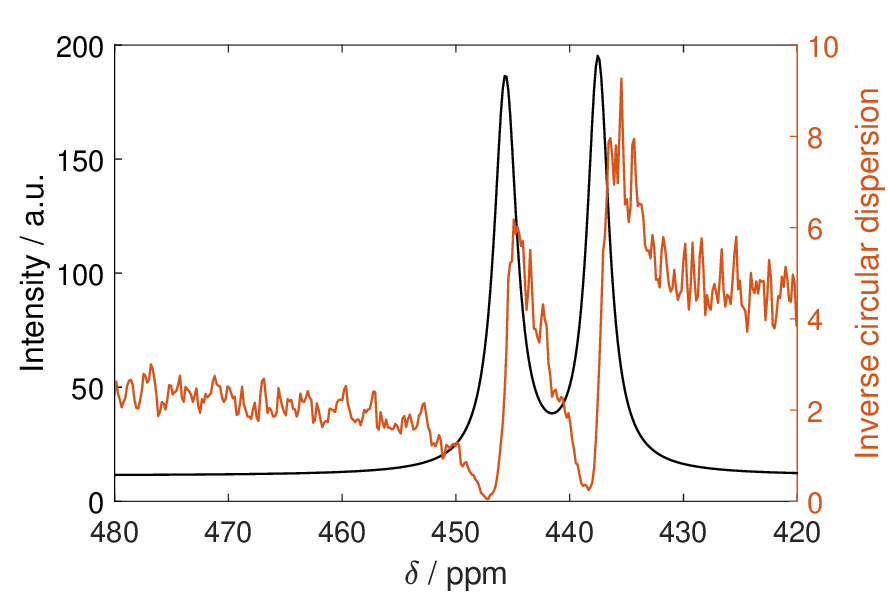}
        \caption{Detail: 480 to 420 ppm}
        \label{fig:fig4b}
    \end{subfigure}
    \caption{Failure of the phasing due to the background signal.}
    \label{fig:fig4}
\end{figure}

It is apparent that, with the application of mFTT, the spectrum is reconstructed with no phase alteration, at variance with the outcome of FT (Figure \ref{fig:fig1}.\subref{fig:fig1d}). It can be noted that the reconstructed spectrum displays only a minor lineshape distortion in the most shifted peaks (Figure \ref{fig:fig3}.\subref{fig:fig3b} and, at variance with the standard phasing approach (Figure \ref{fig:fig1}.\subref{fig:fig1e}) the spectrum has no baseline distortion.\\
As already noted in the FTT paper\cite{fukazawa2011post}, the quantitative information is unfortunately lost. This precludes a quantitative application, but has no impact when the task is simply to observe a signal, as done in \cite{gade2021observability}.\\
As stated above, we have assumed that the probehead is devoid of background signal arising from the components sitting in the inhomogeneous field in the vicinity of the coils. This is usually possible only in dedicated designs \cite{luchinat2001development}. 
The presence of a large background signal impedes the successful application of the mFTT method, because the contribution of the background to the phase of each point in the spectrum can be larger than that of the signal itself. Furthermore, when the signal has a phase opposite to that of the background, the contribution of noise becomes more important. This is clearly shown in figure \ref{fig:fig4}. To model this situation, a gaussian peak centered at 0 ppm, with FWHM 320 ppm and 5 times as intense as the signal from the sample was added to the spectrum. Panel \ref{fig:fig4}.\subref{fig:fig4b} shows the same spectral region as \ref{fig:fig3}.\subref{fig:fig3b}, demonstrating that the mFTT phasing did not succeed as a result of the presence of the background. A possible workaround is discussed in the next section.

\subsection{Experimental Results}
\begin{figure}
    \centering
    \begin{subfigure}[t]{0.70\textwidth}
        \centering
        \includegraphics[width=0.95\linewidth]{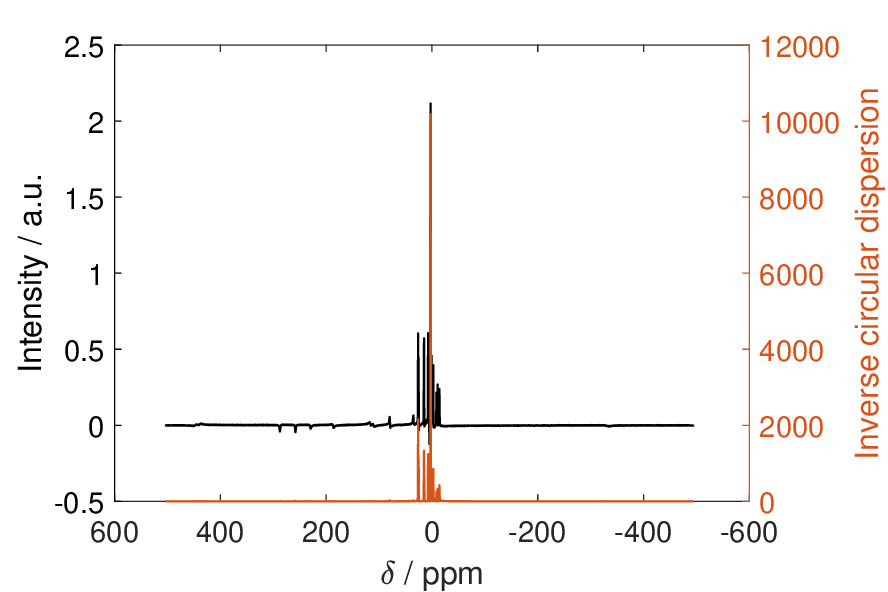} 
        \caption{Full spectrum} \label{fig:fig5a}
    \end{subfigure}

    \begin{subfigure}[t]{0.49\textwidth}
        \centering
        \includegraphics[width=0.95\linewidth]{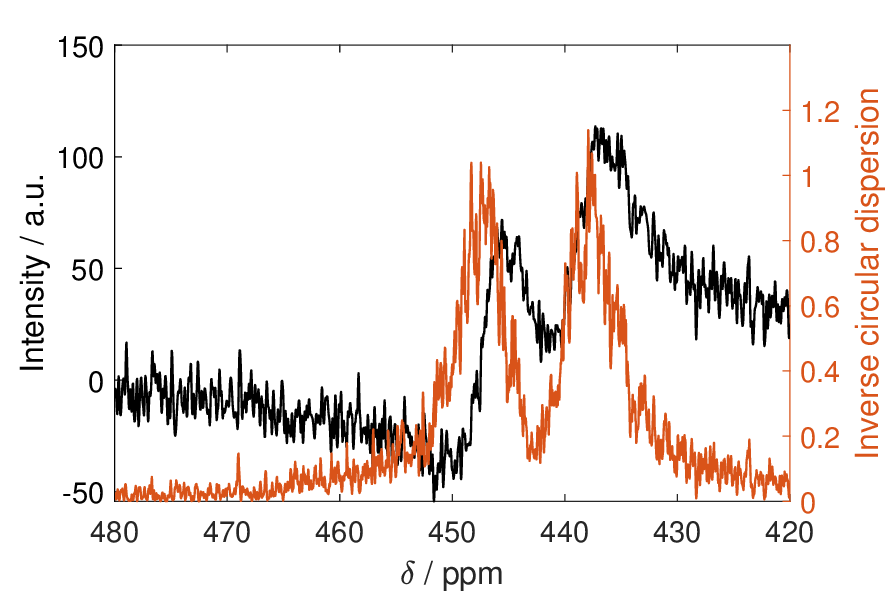} 
        \caption{Detail: 480 to 420 ppm} \label{fig:fig5b}
    \end{subfigure}
    \hfill
    \begin{subfigure}[t]{0.49\textwidth}
        \centering
        \includegraphics[width=0.95\linewidth]{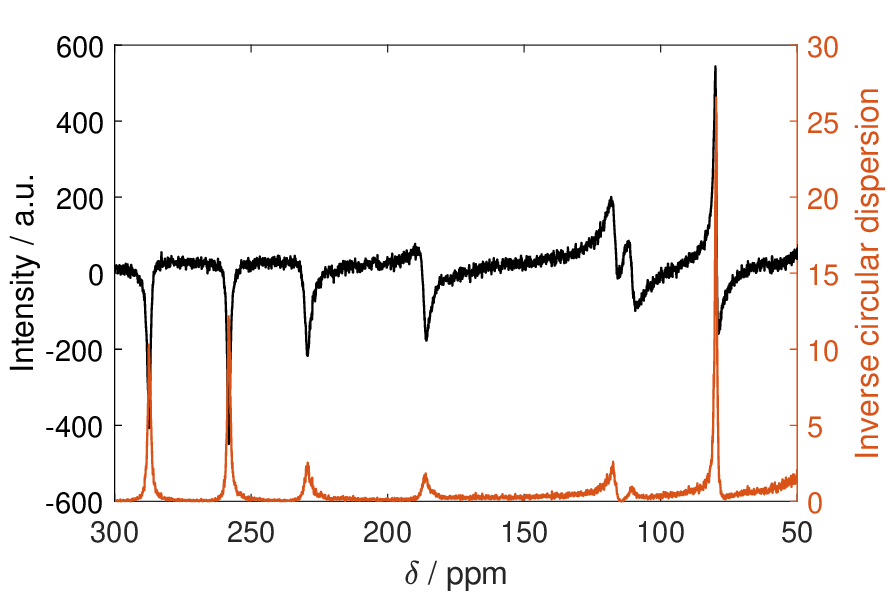} 
        \caption{Detail: 300 to 50 ppm} \label{fig:fig5c}
    \end{subfigure}
    \begin{subfigure}[t]{0.49\textwidth}
        \centering
        \includegraphics[width=0.95\linewidth]{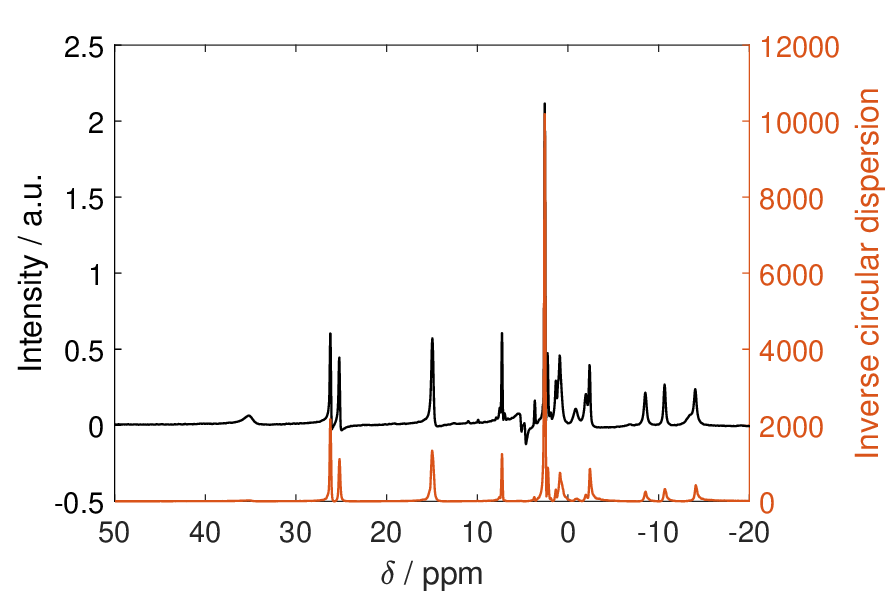} 
        \caption{Detail: 50 to -20 ppm} \label{fig:fig5d}
    \end{subfigure}
    \hfill
    \begin{subfigure}[t]{0.49\textwidth}
        \centering
        \includegraphics[width=0.95\linewidth]{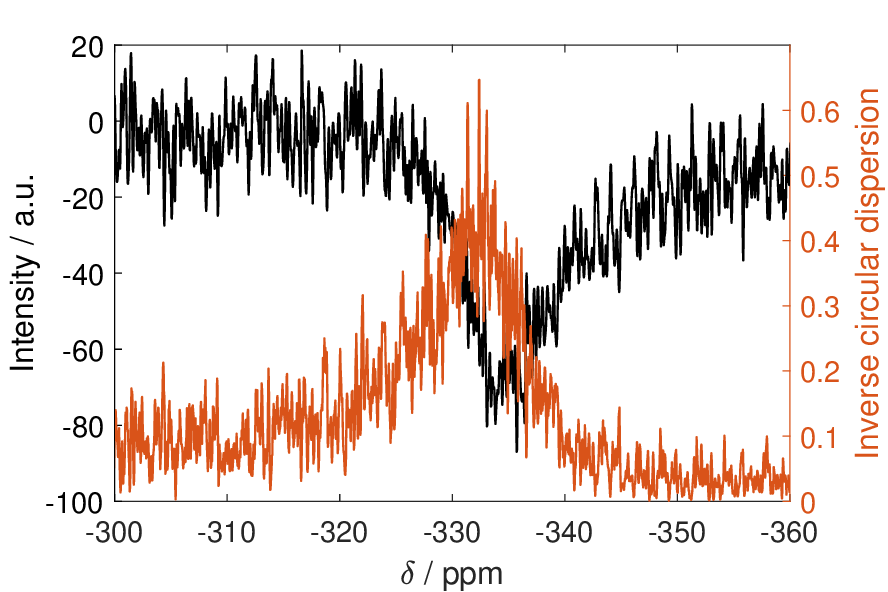} 
        \caption{Detail: -300 to -360 ppm} \label{fig:fig5e}
    \end{subfigure}

    \caption{Experimental spectra of Ni-SAL-HDPT\cite{sacconi1966high} at \SI{9.4}{T} (\SI{400}{MHz} \isotope[1]{H} Larmor frequency) in black, and the reconstructed spectrum devoid of phase distortion in red \subref{fig:fig5a}. Panels \subref{fig:fig5b}-\subref{fig:fig5e} show different details of the full spectrum. An enlargement of panel \subref{fig:fig5d} is shown in figure \ref{fig:figS2}}     \label{fig:fig5}
\end{figure}

\begin{figure}
    \centering
    \begin{subfigure}[t]{0.70\textwidth}
        \centering
        \includegraphics[width=0.95\linewidth]{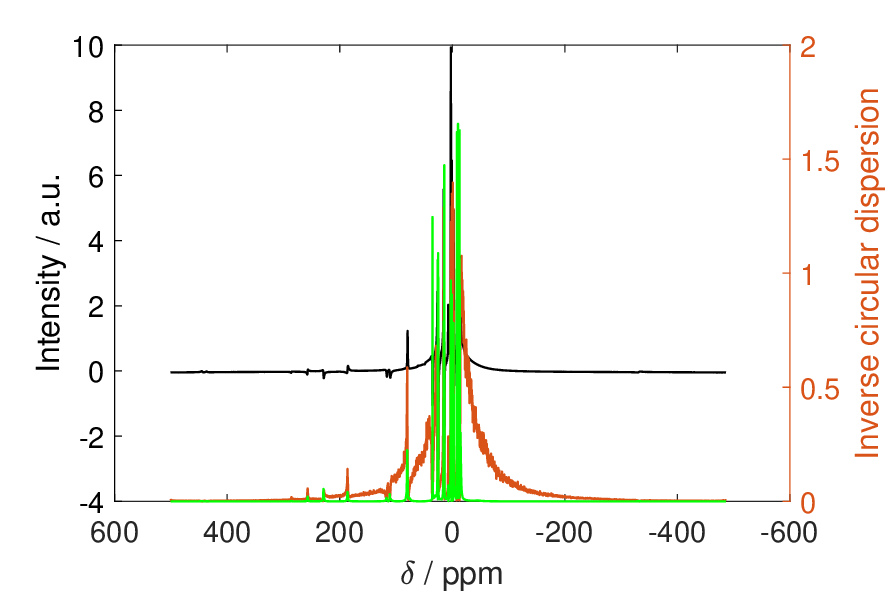} 
        \caption{Full spectrum} \label{fig:fig6a}
    \end{subfigure}

    \begin{subfigure}[t]{0.49\textwidth}
        \centering
        \includegraphics[width=0.95\linewidth]{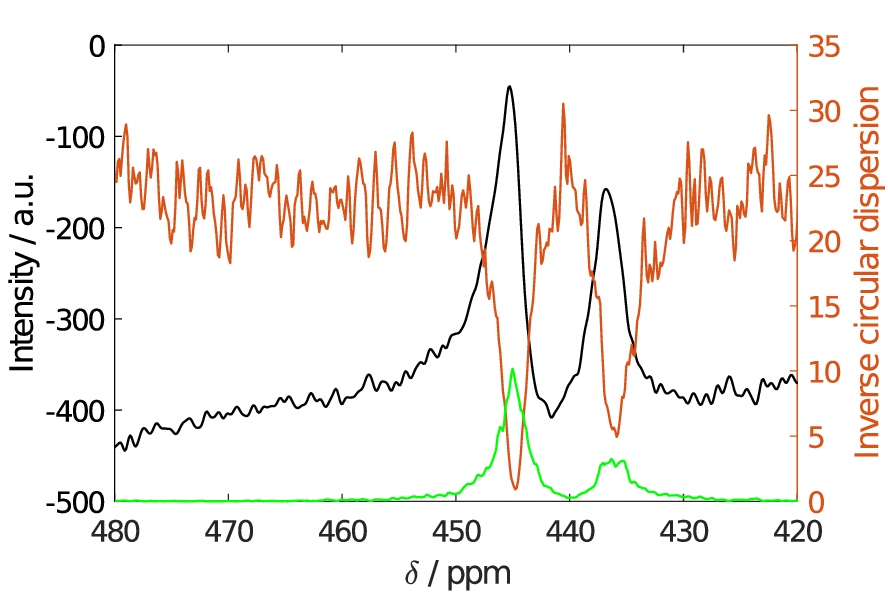} 
        \caption{Detail: 480 to 420 ppm} \label{fig:fig6b}
    \end{subfigure}
    \hfill
    \begin{subfigure}[t]{0.49\textwidth}
        \centering
        \includegraphics[width=0.95\linewidth]{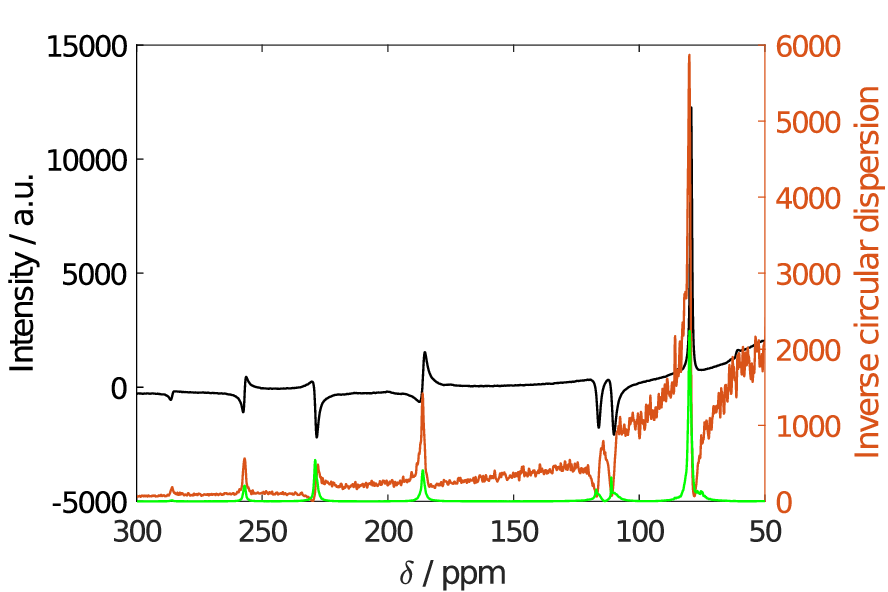} 
        \caption{Detail: 300 to 50 ppm} \label{fig:fig6c}
    \end{subfigure}
    \begin{subfigure}[t]{0.49\textwidth}
        \centering
        \includegraphics[width=0.95\linewidth]{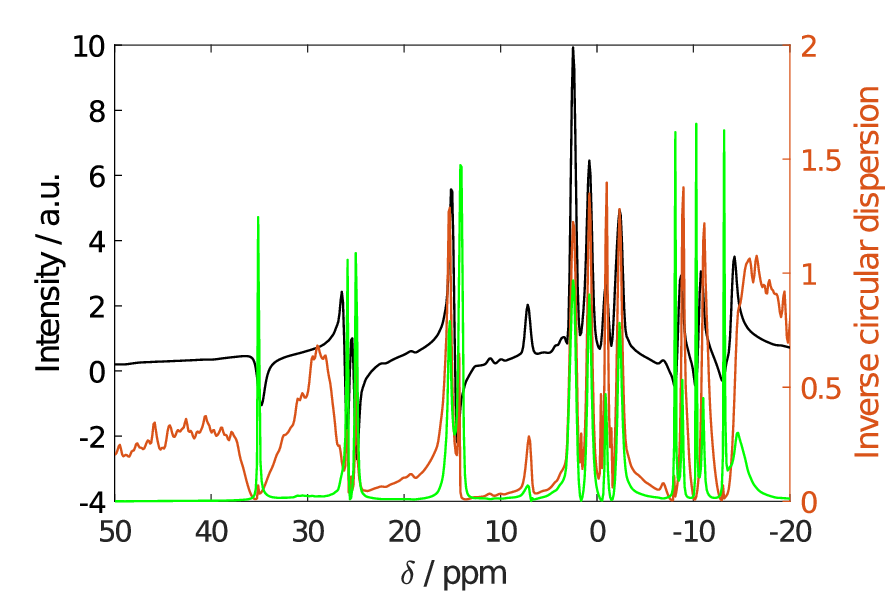} 
        \caption{Detail: 50 to -20 ppm} \label{fig:fig6d}
    \end{subfigure}
    \hfill
    \begin{subfigure}[t]{0.49\textwidth}
        \centering
        \includegraphics[width=0.95\linewidth]{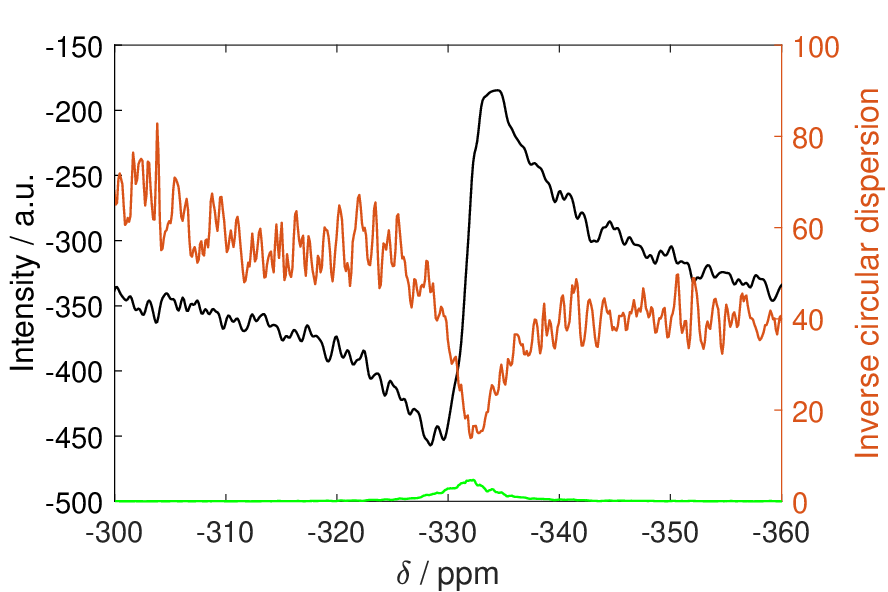} 
        \caption{Detail: -300 to -360 ppm} \label{fig:fig6e}
    \end{subfigure}

    \caption{Experimental spectra of Ni-SAL-HDPT\cite{sacconi1966high} at \SI{22.4}{T} (\SI{950}{MHz} \isotope[1]{H} Larmor frequency) in black, the reconstructed spectrum without background subtraction in red and the reconstructed spectrum with background subtraction in green \subref{fig:fig6a}. Panels \subref{fig:fig6b}-\subref{fig:fig6e} show different details of the full spectrum. Figure \ref{fig:figS3} shows the comparison between the black and green spectra only.}     \label{fig:fig6}
\end{figure}

With the results on the simulated data at hand, it is possible to proceed through the analysis of real datasets, acquired at two different fields. The experiment at \SI{9.4}{T} (see subsection \ref{expdetails}) are acquired with a dedicated probe capable of delivering high-power pulses. Therefore, the acquired spectra suffer from a smaller phase distortion as opposed to the simulated high-field spectra. Furthermore, the probehead has a very limited background, representing an ideal case for testing the method.\\
Figure \ref{fig:fig5} shows the effect of the mFTT reconstruction on the experimental spectra of Ni-SAL-HDPT acquired at \SI{9.4}{T}.
In line with the tests on simulated data, the spectrum appears to be reconstructed with no phase distortion. Also in line with the expectations, the relative intensities of the peaks are altered, with the less-intense peaks being reduced with respect to the more intense ones.\\
These results confirm the validity of the mFTT approach in obtaining a phased spectrum using the statistical analysis of the phase angles across a series of spectra. However, these results are obtained within an ideal playground. How does this method perform in sub-ideal conditions? To check this, the same analysis was performed on the Ni-SAL-HDPT spectra acquired at \SI{22.4}{T}. Given that the probehead as a very broad background, we also proceeded to fit the background in the real and in the imaginary channel separately using a spline interpolant as implemented in MATLAB, with $1.2\cdot10^8$ smoothing parameter. As already noted in subsection \ref{subsec:synthtest}, if the background is not subtracted, the mFTT reconstruction does not succeed. On the contrary, when the background is subtracted, the reconstruction yields a phased spectrum, as shown in figure \ref{fig:fig6}. 

\begin{figure}
    \centering
    \begin{subfigure}[t]{0.70\textwidth}
        \centering
        \includegraphics[width=0.95\linewidth]{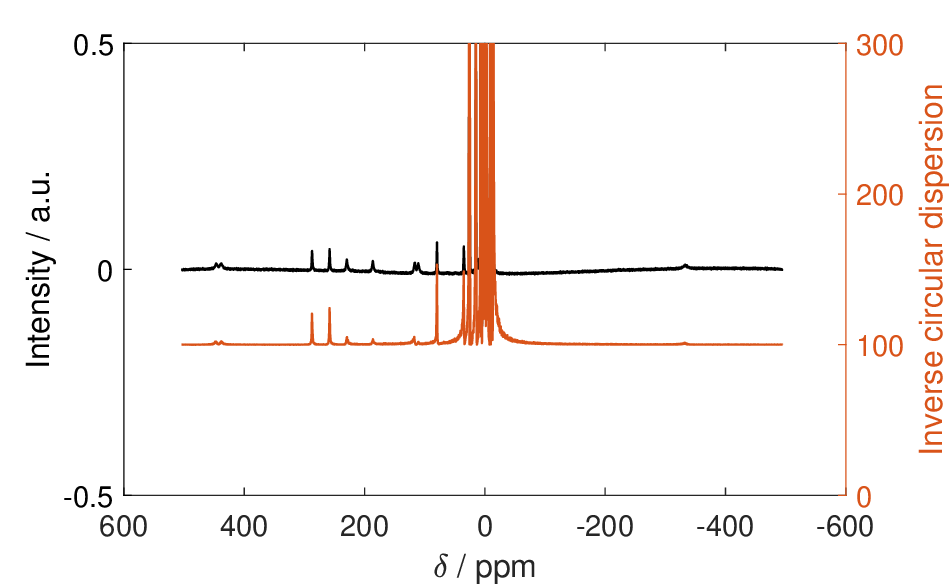} 
        \caption{Full spectrum at \SI{9.4}{T}} \label{fig:fig7a}
    \end{subfigure}

    \begin{subfigure}[t]{0.49\textwidth}
        \centering
        \includegraphics[width=0.95\linewidth]{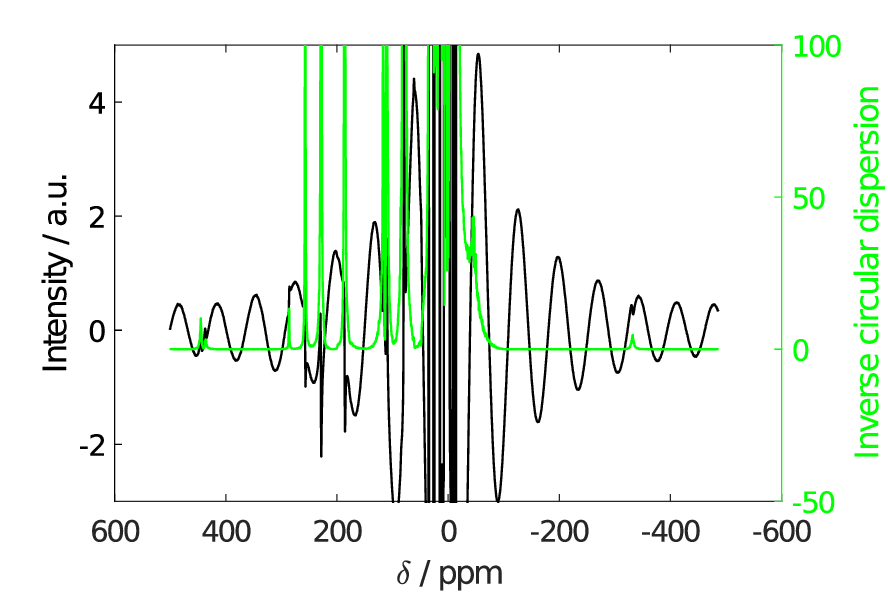} 
        \caption{Full spectrum at \SI{22.4}{T}} \label{fig:fig7b}
    \end{subfigure}
    \hfill
    \begin{subfigure}[t]{0.49\textwidth}
        \centering
        \includegraphics[width=0.95\linewidth]{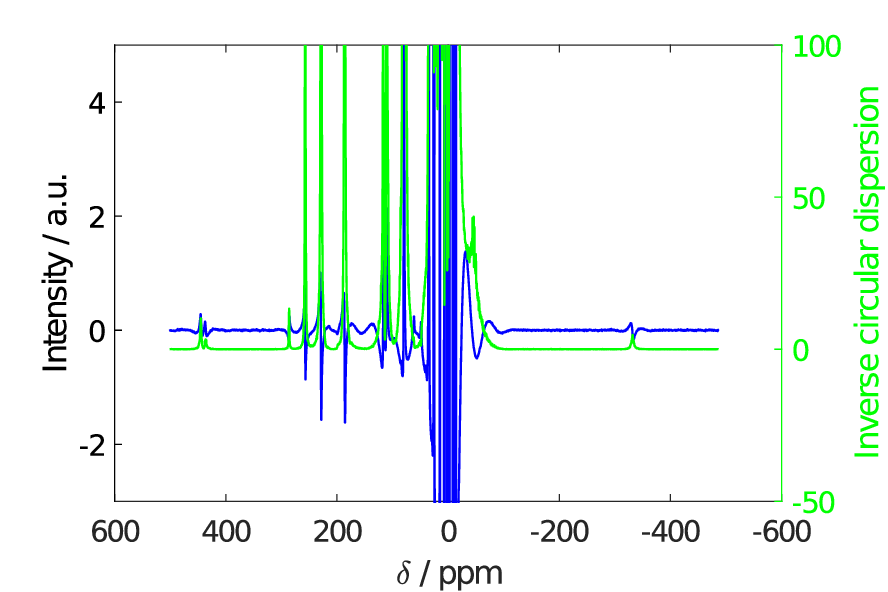} 
        \caption{Full spectrum at \SI{22.4}{T} with subtracted background} \label{fig:fig7c}
    \end{subfigure}

    \caption{Experimental spectra of Ni-SAL-HDPT\cite{sacconi1966high} at \SI{9.4}{T} (\SI{400}{MHz} \isotope[1]{H} Larmor frequency) after manual phasing in black and the reconstructed spectrum in green \subref{fig:fig7a}. Panels \subref{fig:fig7b} and \subref{fig:fig7c} show the comparison between the spectra at \SI{22.4}{T} (\SI{950}{MHz} \isotope[1]{H} Larmor frequency. The reconstructed spectrum (with background subtraction is shown in green and compared to the manually-phased spectra without (black in panel \subref{fig:fig7b}) and with (blue in panel \subref{fig:fig7c}) background subtraction.}\label{fig:fig7}
\end{figure}

The spectra shown in figure \ref{fig:fig6} clearly demonstrate the potential of the mFTT method to achieve uniform phasing across the spectrum without distorting the baseline, even though the data must be background subtracted to achieve a good result.\\
How does this compare to the standard phasing approach? In figure \ref{fig:fig7}, the spectra of \ref{fig:fig5}.\subref{fig:fig5a} and \ref{fig:fig6}.\subref{fig:fig6a} are compared with the manually-phased ones. Note that the spectrum at 950 can only be fased within the first $\pm$ 100 ppm of offset.

\section{Conclusions}
Phasing the NMR spectra of paramagnetic compounds, where the signals are spread over hundreds of ppm, can be challenging also for experienced spectroscopists, and usually requires a tedious process of manual phasing and extensive baseline correction.\\
It is here demonstrated that a method based on the statistical analysis of the phase of each point in a series of spectra acquired on the same sample - which is based on a method proposed earlier by Fukazawa, Takeda and Takegoshi \cite{fukazawa2011post} - is applicable to obtain uniformly-phased pNMR spectra.  This method is, in principle, applicable to all broadband or wideline NMR applications, even though it comes at the price of losing the quantitative information arising from signal intensity. The phasing approach presented in this work is easy to implement, relatively robust - as long as the signals and the background from the probehead and from the electronics can be separated - and yields phased spectra also in cases where manual phasing fails, preventing the spectral manipulation that is usually required and that may cause the loss of some peaks.

\section*{Acknowledgements}
The author wants to thank Claudio Luchinat for his continued mentoring, for encouragement towards maintaining curiosity (and for instructions on the subtleties of pNMR).
Fruitful discussion with Mario Piccioli and Massimo Lucci, and suggestions by Jean-Nicolas Dumez are gratefully acknowledged.
This work has been supported by the Fondazione Cassa di Risparmio di Firenze, the Italian Ministero della Salute through the grant GR-2016-02361586, by the Italian Ministero dell’Istruzione, dell’Università e della Ricerca through the “Progetto Dipartimenti di Eccellenza 2018–2022” to the Department of Chemistry “Ugo Schiff” of the University of Florence, and by the University of Florence through the “Progetti Competitivi per Ricercatori”. The author acknowledges the support and the use of resources of Instruct-ERIC, a landmark ESFRI project, and specifically the CERM/CIRMMP Italy center. The author also acknowledges the H2020 projects iNEXT-Discovery (Grant 871037), TIMB3 (Grant 810856) and HIRES-MULTIDYN (Grant 899683) and Panacea (Grant 101008500).

 \bibliographystyle{elsarticle-num} 
 \bibliography{cas-refs}

\pagebreak

\begin{center}
\textbf{\large Supplemental Materials: Phase distortion-free pNMR spectra}
\end{center}

\setcounter{section}{0}
\setcounter{equation}{0}
\setcounter{figure}{0}
\setcounter{table}{0}
\setcounter{page}{1}
\makeatletter
\renewcommand{\thesection}{S\arabic{section}}
\renewcommand{\theequation}{S\arabic{equation}}
\renewcommand{\thefigure}{S\arabic{figure}}

\renewcommand{\bibnumfmt}[1]{[S#1]}
\renewcommand{\citenumfont}[1]{S#1}

\section{Supplementary Figures}

\begin{figure}[h]
    \centering
        \includegraphics[width=0.95\linewidth]{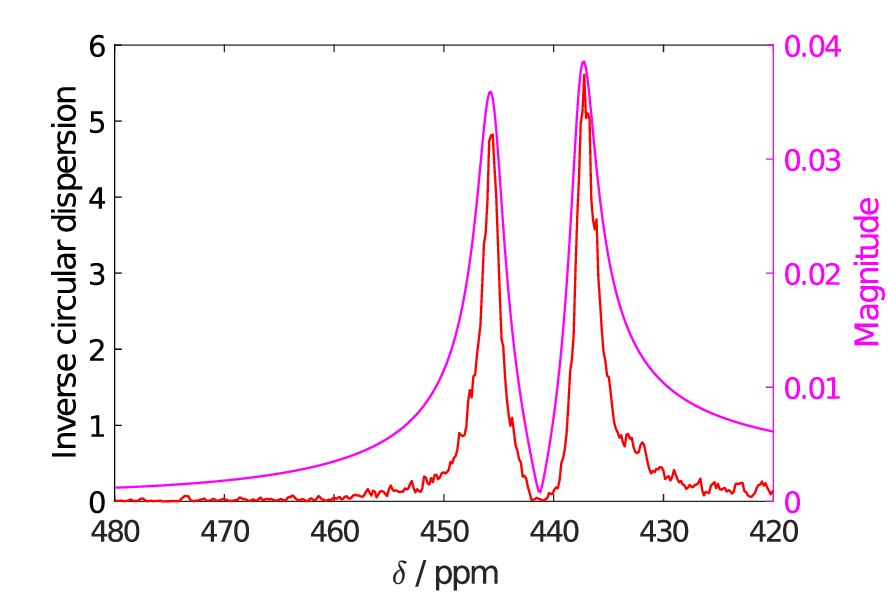} 
        \caption{Lineshape distortion caused by the evaluation of the magnitude (in magenta) as compared to the circular dispersion (in red), in the spectrum shown in panel \subref{fig:fig3b} of figure \ref{fig:fig3}} \label{fig:figSmagn}

\end{figure}

\begin{figure}[h]
    \centering
        \includegraphics[width=0.95\linewidth]{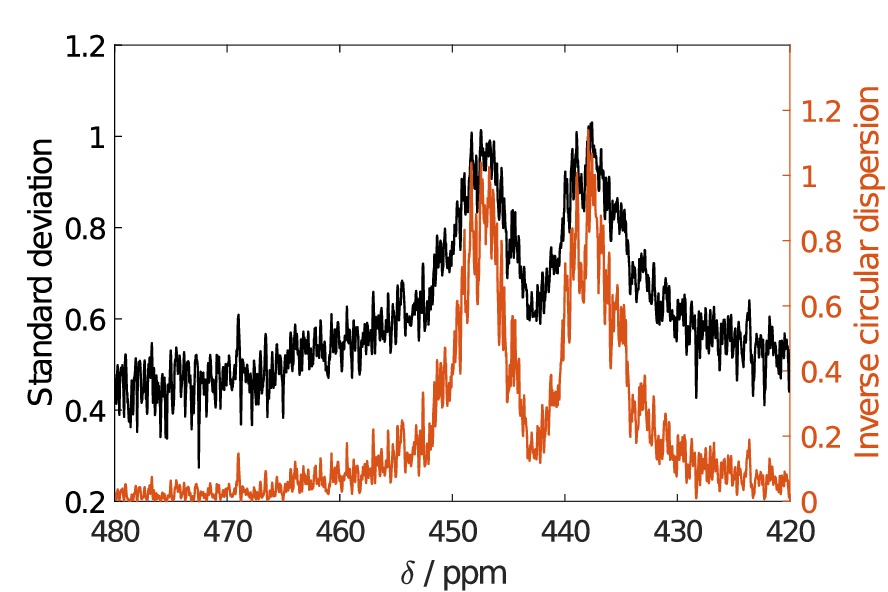} 
        \caption{Lineshape distortion caused by the evaluation of the standard deviation (in black) as compared to the circular dispersion (in red), in the spectrum shown in panel \subref{fig:fig5b} of figure \ref{fig:fig5}} \label{fig:figS1}

\end{figure}

\begin{figure}[h]
    \centering
        \includegraphics[width=0.95\linewidth]{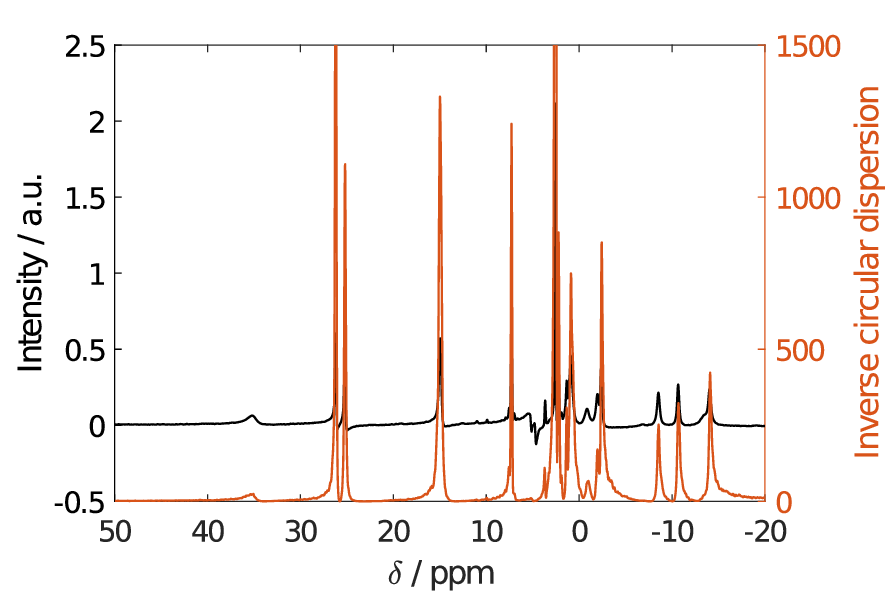} 
        \caption{Enlargement of figure \ref{fig:fig5}, panel \subref{fig:fig5d}} \label{fig:figS2}

\end{figure}

\begin{figure}[h]
    \centering
    \begin{subfigure}[t]{0.70\textwidth}
        \centering
        \includegraphics[width=0.95\linewidth]{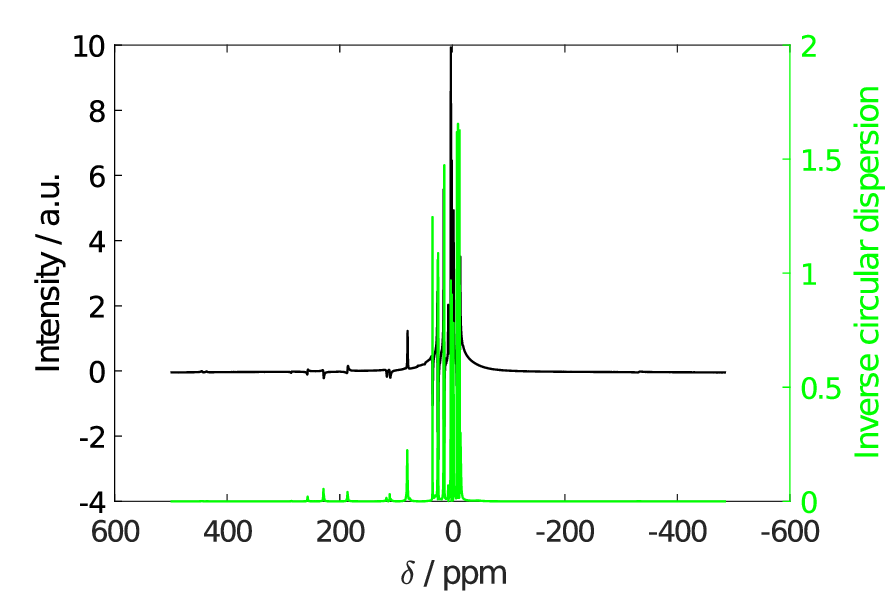} 
        \caption{Full spectrum} \label{fig:figS3a}
    \end{subfigure}

    \begin{subfigure}[t]{0.49\textwidth}
        \centering
        \includegraphics[width=0.95\linewidth]{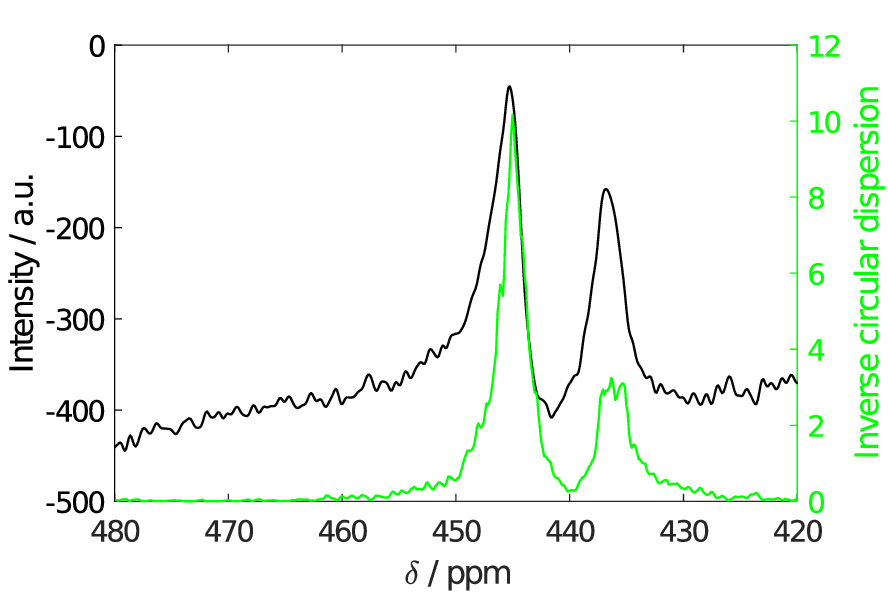} 
        \caption{Detail: 480 to 420 ppm} \label{fig:figS3b}
    \end{subfigure}
    \hfill
    \begin{subfigure}[t]{0.49\textwidth}
        \centering
        \includegraphics[width=0.95\linewidth]{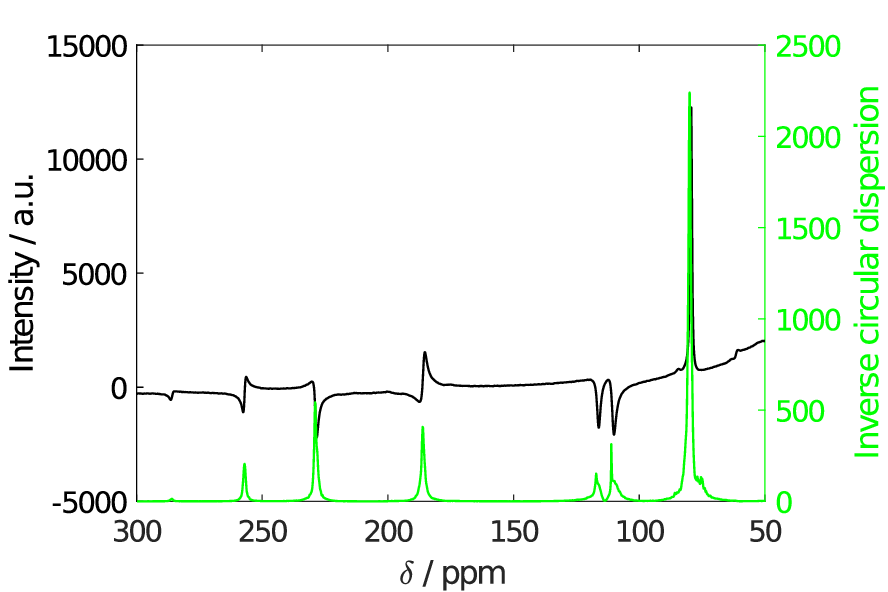} 
        \caption{Detail: 300 to 50 ppm} \label{fig:figS3c}
    \end{subfigure}
    \begin{subfigure}[t]{0.49\textwidth}
        \centering
        \includegraphics[width=0.95\linewidth]{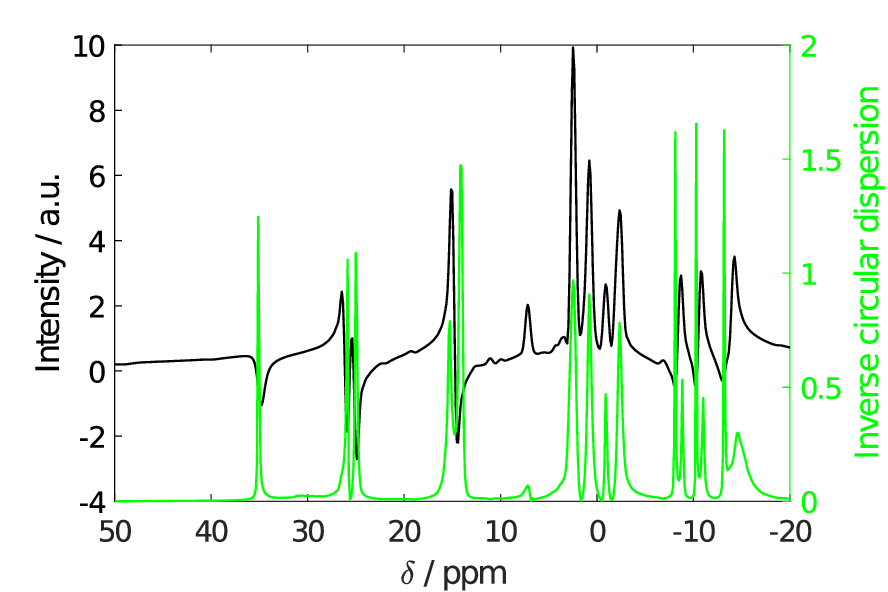} 
        \caption{Detail: 50 to -20 ppm} \label{fig:figS3d}
    \end{subfigure}
    \hfill
    \begin{subfigure}[t]{0.49\textwidth}
        \centering
        \includegraphics[width=0.95\linewidth]{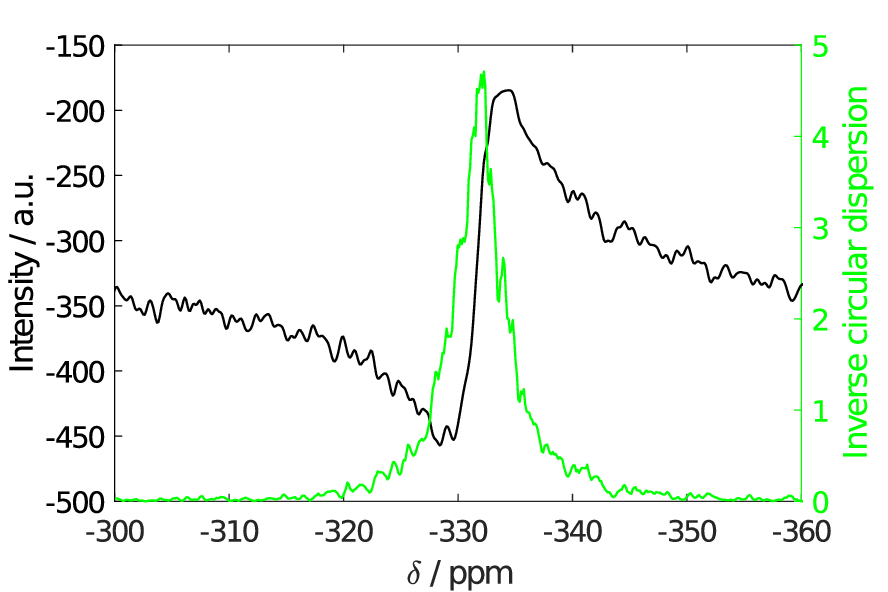} 
        \caption{Detail: -300 to -360 ppm} \label{fig:figS3e}
    \end{subfigure}

    \caption{Experimental spectra of Ni-SAL-HDPT\cite{sacconi1966high} at \SI{22.4}{T} (\SI{950}{MHz} \isotope[1]{H} Larmor frequency) in black, and the reconstructed spectrum with background subtraction in green \subref{fig:figS3a}. Panels \subref{fig:figS3b}-\subref{fig:figS3e} show different details of the full spectrum.}     \label{fig:figS3}
\end{figure}

\section{Code for performing mFTT}
A sample code generating the test on the simulated data is provided below \ref{lst:simutest}. It requires the packages CircStat2012a\\ (http://philippberens.wordpress.com/code/circstats/), applying the modifications indicated in listing \ref{lst:circstat} to the functions circ\_r and circ\_std.
To run the reconstruction on a real experimental dataset, the steps necessary for implementation on a bruker instrument are listed here:
\begin{enumerate}
\item modify the sequence to store the FID as a different increment every nth-scan
\item process the spectra in the direct dimension (xf2) applying "quad" or "qpol" as required by the possible misset in the real and imaginary channels
\item import the spectrum using the brukerimport function from the GNAT library \\(https://www.nmr.chemistry.manchester.ac.uk/?q=node/430)\\ with the modifications indicated in listing \ref{lst:brukerimport}
\item apply the functions indicated in listing \ref{lst:mFTT} 
\end{enumerate}

\begin{lstlisting}[caption={Code for running mFFT on simulated data \label{lst:simutest}}
style=Matlab-editor,
basicstyle=\mlttfamily
]
clear all
%% generate the fid 
B0 = 22.4;
shifts = [-13.969545574250091; 25.18594197457872; -2.3479228587718532; 14.896409910147531; 445.40669790036816; 116.39006657698943; 228.9256609604365; 110.75630475732792; 185.9769009045458; -13.498722161309246; 26.158175424997744; 2.5155746331251922; 14.964343443187055; 437.2614207347274; -1.9933524248824386; -0.9102327021243197; 79.42224147320199; 286.39155614543347; 257.5951336973578; 34.64304849952982; -8.58913008234927; -10.603001316056927; -332.75478518844994];
r2 = [190.35040780290768; 100; 100; 100;7190.55428662293;2395.710048502666;2021.9752357543648;2814.9171953328796;1400.1773322048866;186.35061096093486;100;100;100;6855.207755624197;234.8788023434185;452.3995796901293;2697.887036804826;3845.978076777599;3053.0846049886845;1996.598779127256;390.118174091022;251.3605204271949;7550.075097212369];
Larmor = B0*42.5;
peaks = [shifts r2 r2]; %assume that R1 = R2
npeaks = size(peaks,1);
t90 = 7;
taup = 1;
omeganut = 2e6*pi/(4*t90);
omegas = zeros(npeaks,2);
angles = omegas;
am = pi*taup/(2*t90);
phase = zeros(npeaks,1);
intensity = phase;
td = 4096;
dw = 8e-7;
d1 = 0.02;
de = 8e-6;
aq = linspace(de,td*dw+de,td);
aqperf = linspace(0,td*dw,td);
taurec = max(aq)+3e-5+d1;
fid0 = 1j.*zeros(1,td);
fid0perf = fid0;
fid0nophbutde = fid0;
fid0nodebutph = fid0;
for i=1:npeaks
    omegas(i,1) = 2*pi*peaks(i,1)*Larmor;
    omegas(i,2) = (omeganut^2+omegas(i,1)^2)^0.5;
    Delta = abs(omegas(i,1))/omeganut;
    length = (1+Delta^2)^0.5;
    theta = atan(1/Delta);
    alpha = am*length;
    angles(i,1) = theta;
    angles(i,2) = alpha;
    phase(i) = atan(cos(theta)*(1-cos(alpha))/sin(alpha));
    intensity(i) = sin(theta)*(((sin(alpha))^2+(cos(theta)*(1-cos(alpha)))^2)^0.5);
    temp = exp(1j*omegas(i,1).*aq).*exp(-peaks(i,2).*aq);
    fid0 = fid0 + intensity(i)*exp(1j*phase(i))*(1-exp(-peaks(i,3)*taurec)).*temp;
    fid0nophbutde = fid0nophbutde + (1-exp(-peaks(i,3)*taurec)).*temp;    
    temp = exp(1j*omegas(i,1).*aqperf).*exp(-peaks(i,2).*aqperf);
    fid0perf = fid0perf + (1-exp(-peaks(i,3)*taurec)).*temp;
    fid0nodebutph = fid0nodebutph + intensity(i)*exp(1j*phase(i))*(1-exp(-peaks(i,3)*taurec)).*temp;

end
ssb = 2;
lb = 10000;
qsin = sin((pi-pi/ssb)*((aqperf)./max(aqperf))+(pi/ssb).*ones(1,td)).^2;
fid = [fid0.*qsin 1j.*zeros(1,td)];
normfid = norm(fid0);
ppmscale = linspace(1/(2*dw),-1/(2*dw),td*2)./Larmor;
st = fliplr(fftshift(fft(fid)));
st = st./max(st);
%% spectrum with all distortions
figure(1)
plot(ppmscale,real(st),'color','black','LineWidth',1);
set(gca, 'XDir','reverse')
xlabel('\delta / ppm', 'FontSize', 16) 
ylabel('Intensity / a.u.', 'FontSize', 16) 
butto = get(gca,'XTickLabel');
set(gca,'XTickLabel',butto,'fontsize',14)
butto = get(gca,'YTickLabel');
set(gca,'YTickLabel',butto,'fontsize',14)
set(gcf, 'Units', 'centimeters', 'Position', [0, 0, 15, 10], 'PaperUnits', 'centimeters', 'PaperSize', [15, 10])
saveas(gcf,'fig1d.eps')
%% pure spectrum
figure(2)
s2 = fliplr(fftshift(fft([fid0perf.*qsin 1j.*zeros(1,td)])));
plot(ppmscale,real(s2.*exp(1j*0.15)./max(s2)),'color','black','LineWidth',1);
set(gca, 'XDir','reverse')
set(gca, 'XDir','reverse')
xlabel('\delta / ppm', 'FontSize', 16) 
ylabel('Intensity / a.u.', 'FontSize', 16) 
butto = get(gca,'XTickLabel');
set(gca,'XTickLabel',butto,'fontsize',14)
butto = get(gca,'YTickLabel');
set(gca,'YTickLabel',butto,'fontsize',14)
set(gcf, 'Units', 'centimeters', 'Position', [0, 0, 15, 10], 'PaperUnits', 'centimeters', 'PaperSize', [15, 10])
saveas(gcf,'fig1a.eps')
%% only dead time
figure(3)
s3 = fliplr(fftshift(fft([fid0nophbutde.*qsin 1j.*zeros(1,td)])));
plot(ppmscale,real(s3./max(s3)),'color','black','LineWidth',1);
set(gca, 'XDir','reverse')
set(gca, 'XDir','reverse')
xlabel('\delta / ppm', 'FontSize', 16) 
ylabel('Intensity / a.u.', 'FontSize', 16) 
butto = get(gca,'XTickLabel');
set(gca,'XTickLabel',butto,'fontsize',14)
butto = get(gca,'YTickLabel');
set(gca,'YTickLabel',butto,'fontsize',14)
set(gcf, 'Units', 'centimeters', 'Position', [0, 0, 15, 10], 'PaperUnits', 'centimeters', 'PaperSize', [15, 10])
saveas(gcf,'fig1c.eps')
%% only pulse imperfection
figure(4)
s4 = fliplr(fftshift(fft([fid0nodebutph.*qsin 1j.*zeros(1,td)])));
plot(ppmscale,real(s4./max(s4)),'color','black','LineWidth',1);
set(gca, 'XDir','reverse')
set(gca, 'XDir','reverse')
xlabel('\delta / ppm', 'FontSize', 16) 
ylabel('Intensity / a.u.', 'FontSize', 16) 
butto = get(gca,'XTickLabel');
set(gca,'XTickLabel',butto,'fontsize',14)
butto = get(gca,'YTickLabel');
set(gca,'YTickLabel',butto,'fontsize',14)
set(gcf, 'Units', 'centimeters', 'Position', [0, 0, 15, 10], 'PaperUnits', 'centimeters', 'PaperSize', [15, 10])
saveas(gcf,'fig1b.eps')
%% prova a
a = 4.81818;
b = 32.5;
c = -0.507988;
figure (5)
anglecomp = (a + b.*(linspace(-1,1,td*2)+c));
phasecomp = exp(1j.*anglecomp);
plot(ppmscale,real(st.*phasecomp),'color','black','LineWidth',1);
xlabel('\delta / ppm', 'FontSize', 16) 
ylabel('Intensity / a.u.', 'FontSize', 16) 
butto = get(gca,'XTickLabel');
set(gca,'XTickLabel',butto,'fontsize',14)
butto = get(gca,'YTickLabel');
set(gca,'YTickLabel',butto,'fontsize',14)
hold on
yyaxis right
plot(ppmscale,anglecomp - 2*pi*floor( (anglecomp+pi)/(2*pi) ),'LineWidth',1);
ylabel('Phase compensation / rad', 'FontSize', 16) 
set(gca, 'XDir','reverse')
butto = get(gca,'YTickLabel');
set(gca,'YTickLabel',butto,'fontsize',14)
set(gcf, 'Units', 'centimeters', 'Position', [0, 0, 15, 10], 'PaperUnits', 'centimeters', 'PaperSize', [15, 10])
saveas(gcf,'fig1e.eps')

%% generate perturbations
td1eff = 256;
A = fid0'./normfid + normrnd(0+0j,0.01,td,td1eff);
B = [A.*qsin'; 1j.*zeros(td,td1eff)];

stt = fftshift(fft(B));
mcst = abs(st);
ph = angle(stt); 
[angdev, stdph, delta]=circ_std(ph, [], [], 2);



%% plot the histograms

histogram(ph(2936,:),linspace(-pi,pi,90))
xlabel('phase / rad', 'FontSize', 16) 
ylabel('count ', 'FontSize', 16) 
butto = get(gca,'XTickLabel');
set(gca,'XTickLabel',butto,'fontsize',14)
butto = get(gca,'YTickLabel');
set(gca,'YTickLabel',butto,'fontsize',14)
set(gcf, 'Units', 'centimeters', 'Position', [0, 0, 15, 10], 'PaperUnits', 'centimeters', 'PaperSize', [15, 10])
saveas(gcf,'fig2a.eps')
histogram(ph(1316,:),linspace(-pi,pi,90))
xlabel('phase / rad', 'FontSize', 16) 
ylabel('count ', 'FontSize', 16) 
butto = get(gca,'XTickLabel');
set(gca,'XTickLabel',butto,'fontsize',14)
butto = get(gca,'YTickLabel');
set(gca,'YTickLabel',butto,'fontsize',14)
set(gcf, 'Units', 'centimeters', 'Position', [0, 0, 15, 10], 'PaperUnits', 'centimeters', 'PaperSize', [15, 10])
saveas(gcf,'fig2b.eps')
histogram(ph(3150,:),linspace(-pi,pi,90))
xlabel('phase / rad', 'FontSize', 16) 
ylabel('count ', 'FontSize', 16) 
butto = get(gca,'XTickLabel');
set(gca,'XTickLabel',butto,'fontsize',14)
butto = get(gca,'YTickLabel');
set(gca,'YTickLabel',butto,'fontsize',14)
set(gcf, 'Units', 'centimeters', 'Position', [0, 0, 15, 10], 'PaperUnits', 'centimeters', 'PaperSize', [15, 10])
saveas(gcf,'fig2c.eps')

%% make the figures for all the parts of the spectrum
figure(11)
plot(ppmscale,real(s2),'color','black','LineWidth',1);
xlabel('\delta / ppm', 'FontSize', 16) 
ylabel('Intensity / a.u.', 'FontSize', 16) 
butto = get(gca,'XTickLabel');
set(gca,'XTickLabel',butto,'fontsize',14)
butto = get(gca,'YTickLabel');
set(gca,'YTickLabel',butto,'fontsize',14)
hold on
yyaxis right
plot(ppmscale,1./delta,'LineWidth',1);
ylabel('Inverse circular dispersion', 'FontSize', 16) 
set(gca, 'XDir','reverse')
butto = get(gca,'YTickLabel');
set(gca,'YTickLabel',butto,'fontsize',14)
set(gcf, 'Units', 'centimeters', 'Position', [0, 0, 15, 10], 'PaperUnits', 'centimeters', 'PaperSize', [15, 10])
saveas(gcf,'fig3a.eps')

figure(12)
plot(ppmscale,real(s2),'color','black','LineWidth',1);
xlim(gca,[-360 -300])
xlabel('\delta / ppm', 'FontSize', 16) 
ylabel('Intensity / a.u.', 'FontSize', 16) 
butto = get(gca,'XTickLabel');
set(gca,'XTickLabel',butto,'fontsize',14)
butto = get(gca,'YTickLabel');
set(gca,'YTickLabel',butto,'fontsize',14)
hold on
yyaxis right
plot(ppmscale,1./delta,'LineWidth',1);
ylabel('Inverse circular dispersion', 'FontSize', 16) 
set(gca, 'XDir','reverse')
butto = get(gca,'YTickLabel');
set(gca,'YTickLabel',butto,'fontsize',14)
set(gcf, 'Units', 'centimeters', 'Position', [0, 0, 15, 10], 'PaperUnits', 'centimeters', 'PaperSize', [15, 10])
saveas(gcf,'fig3b.eps')

figure(13)
plot(ppmscale,real(s2),'color','black','LineWidth',1);
xlim(gca,[-20 50])
xlabel('\delta / ppm', 'FontSize', 16) 
ylabel('Intensity / a.u.', 'FontSize', 16) 
butto = get(gca,'XTickLabel');
set(gca,'XTickLabel',butto,'fontsize',14)
butto = get(gca,'YTickLabel');
set(gca,'YTickLabel',butto,'fontsize',14)
hold on
yyaxis right
plot(ppmscale,1./delta,'LineWidth',1);
ylabel('Inverse circular dispersion', 'FontSize', 16) 
set(gca, 'XDir','reverse')
butto = get(gca,'YTickLabel');
set(gca,'YTickLabel',butto,'fontsize',14)
set(gcf, 'Units', 'centimeters', 'Position', [0, 0, 15, 10], 'PaperUnits', 'centimeters', 'PaperSize', [15, 10])
saveas(gcf,'fig3c.eps')

figure(14)
plot(ppmscale,real(s2),'color','black','LineWidth',1);
xlim(gca,[50 300])
xlabel('\delta / ppm', 'FontSize', 16) 
ylabel('Intensity / a.u.', 'FontSize', 16) 
butto = get(gca,'XTickLabel');
set(gca,'XTickLabel',butto,'fontsize',14)
butto = get(gca,'YTickLabel');
set(gca,'YTickLabel',butto,'fontsize',14)
hold on
yyaxis right
plot(ppmscale,1./delta,'LineWidth',1);
ylabel('Inverse circular dispersion', 'FontSize', 16) 
set(gca, 'XDir','reverse')
butto = get(gca,'YTickLabel');
set(gca,'YTickLabel',butto,'fontsize',14)
set(gcf, 'Units', 'centimeters', 'Position', [0, 0, 15, 10], 'PaperUnits', 'centimeters', 'PaperSize', [15, 10])
saveas(gcf,'fig3d.eps')

figure(15)
plot(ppmscale,real(s2),'color','black','LineWidth',1);
xlim(gca,[420 480])
xlabel('\delta / ppm', 'FontSize', 16) 
ylabel('Intensity / a.u.', 'FontSize', 16) 
butto = get(gca,'XTickLabel');
set(gca,'XTickLabel',butto,'fontsize',14)
butto = get(gca,'YTickLabel');
set(gca,'YTickLabel',butto,'fontsize',14)
hold on
yyaxis right
plot(ppmscale,1./delta,'LineWidth',1);
ylabel('Inverse circular dispersion', 'FontSize', 16) 
set(gca, 'XDir','reverse')
butto = get(gca,'YTickLabel');
set(gca,'YTickLabel',butto,'fontsize',14)
set(gcf, 'Units', 'centimeters', 'Position', [0, 0, 15, 10], 'PaperUnits', 'centimeters', 'PaperSize', [15, 10])
saveas(gcf,'fig3e.eps')

%% add background to the data

a1 =        5./size(stt,1);  
b1 =     0.01059;  
c1 =      2000; 
x = linspace(0,1,size(stt,1))';
f =  fftshift(fft(a1*(1/(2/(c1^2))^0.5)*exp(-((c1.*x).^2)./(4+1j*b1.*x))));

sttb = stt + f;

ph = angle(sttb); 
angmean = circ_mean(ph, [], 2);

[angdev, stdph, delta]=circ_std(ph, [], [], 2);


figure(16)
plot(ppmscale,real(s2),'color','black','LineWidth',1);
xlabel('\delta / ppm', 'FontSize', 16) 
ylabel('Intensity / a.u.', 'FontSize', 16) 
butto = get(gca,'XTickLabel');
set(gca,'XTickLabel',butto,'fontsize',14)
butto = get(gca,'YTickLabel');
set(gca,'YTickLabel',butto,'fontsize',14)
hold on
yyaxis right
plot(ppmscale,1./delta,'LineWidth',1);
ylabel('Inverse circular dispersion', 'FontSize', 16) 
set(gca, 'XDir','reverse')
butto = get(gca,'YTickLabel');
set(gca,'YTickLabel',butto,'fontsize',14)
set(gcf, 'Units', 'centimeters', 'Position', [0, 0, 15, 10], 'PaperUnits', 'centimeters', 'PaperSize', [15, 10])
saveas(gcf,'fig4a.eps')
figure(15)
plot(ppmscale,real(s2),'color','black','LineWidth',1);
xlim(gca,[420 480])
xlabel('\delta / ppm', 'FontSize', 16) 
ylabel('Intensity / a.u.', 'FontSize', 16) 
butto = get(gca,'XTickLabel');
set(gca,'XTickLabel',butto,'fontsize',14)
butto = get(gca,'YTickLabel');
set(gca,'YTickLabel',butto,'fontsize',14)
hold on
yyaxis right
plot(ppmscale,1./delta,'LineWidth',1);
ylabel('Inverse circular dispersion', 'FontSize', 16) 
set(gca, 'XDir','reverse')
butto = get(gca,'YTickLabel');
set(gca,'YTickLabel',butto,'fontsize',14)
set(gcf, 'Units', 'centimeters', 'Position', [0, 0, 15, 10], 'PaperUnits', 'centimeters', 'PaperSize', [15, 10])
saveas(gcf,'fig4b.eps')

\end{lstlisting}

\begin{lstlisting}[caption={Modifications required in circ\_std \label{lst:circstat}}
style=Matlab-editor,
basicstyle=\mlttfamily
]
% in circ_r
% replace line 1 with 
function [r r2] = circ_r(alpha, w, d, dim)
% after line 51
r2 = sum(w.*exp(2.*1j*alpha),dim);
% after line 54
r2 = abs(r2)./sum(w,dim);
% in circ_std
% replace line 1 with 
function [s s0 delta] = circ_std(alpha, w, d, dim)
% replace line 51 with
[r  r2] = circ_r(alpha,w,d,dim);
%add as last line
delta = (1-r2)./(2*r.^2);

\end{lstlisting}

\begin{lstlisting}[caption={Modifications required in brukerimport \label{lst:brukerimport}}
style=Matlab-editor,
basicstyle=\mlttfamily
]
ProcFIDpathimg=[ProcPathRoot '2ii'];       %after line 1255
fileid_ProcFIDimg=fopen(ProcFIDpathimg ,'r',byte_format_proc); %after line 1258
% this block after line 1284
imagfid=fread(fileid_ProcFIDimg,brukerdata.np*2*SI2,byte_size);
imagfid=imagfid*2^str2num(brukerdata.proc2s.NC_proc);      
imagfid = reshape(...
permute(...
reshape(...
permute(...
reshape(imagfid,XDIM1,XDIM2,NoSM),...
[2 1 3]),...
XDIM2,SI1,NoSM2),...
[2 1 3]),...
SI1,SI2)';
imagfid=imagfid';
% replace line 1287 with 
brukerdata.X2DSPEC=realfid+1j.*imagfid;

\end{lstlisting}

\begin{lstlisting}[caption={Commands for performing mFTT \label{lst:mFTT}}
style=Matlab-editor,
basicstyle=\mlttfamily
]
brukerdata1 = brukerimportimg(1,'nisalphasecorr/12/pdata/1'); %here the name of your dataset
stt = brukerdata1.X2DSPEC;
o1p = str2num(convertCharsToStrings(brukerdata1.acqus.O1))/brukerdata1.sfrq;
sw = brukerdata1.sw;
SI = brukerdata1.SI;
ppmscale = linspace(o1p+sw/2,o1p-sw/2,SI);
s2 = mean(stt,2);
ph = angle(stt); %phase
[angdev, stdph, delta]=circ_std(ph, [], [], 2);
plot(1./delta) %this is the reconstructed spectrum
\end{lstlisting}

\end{document}